 \documentclass [12pt,a4paper,     ]{report} 

\usepackage{graphics}
\usepackage{times}

\DeclareFontFamily{OT1}{times}{}
\DeclareFontShape {OT1}{times}{m }{n }{ <-> ptmr }{}
\DeclareFontShape {OT1}{times}{bx}{n }{ <-> ptmb }{}
\DeclareFontShape {OT1}{times}{m }{it}{ <-> ptmri}{}
\DeclareFontShape {OT1}{times}{bx}{it}{ <-> ptmbi}{}
\usepackage{amsmath}
\usepackage{amsfonts}
\usepackage{amssymb}
\usepackage{latexsym}
\usepackage[german,english]{babel}
\newcommand{\cl}{C \kern -0.1em \ell_\pi} 

\newcommand{\CON}{\overline}          

\newcommand{\Scal}{\mathbb{S}}        

\newcommand{\VEC}{\vec{\kern +.1em[}} 
\newcommand{\TOR}{\vec{\kern +.2em]}} 
\newcommand{\BRA}{\langle\kern -.2em\langle} 
\newcommand{\KET}{\rangle\kern -.2em\rangle} 

\newcommand{\REA}{\operatorname{Re}}  

\newcommand{\Q}{[\hspace{1.mm}]} 
\newcommand{\A}{(\hspace{.5mm})} 

\begin{document}

\renewcommand{\contentsname}  {Inhaltsverzeichnis}
\renewcommand{\listfigurename}{Abbildungsverzeichnis}
\renewcommand{\chaptername}   {Kapitel}
\renewcommand{\figurename}    {Figur}

\title{\bf  \vspace{-3cm}  {\huge DIE\\
                       FUNKTIONENTHEORETISCHEN
                   BEZIEHUNGEN DER MAXWELLSCHEN
                         AETHERGLEICHUNGEN} \\
                             ~\\
          {\large EIN BEITRAG ZUR RELATIVIT\"{A}TS-\\
                     UND ELEKTRONENTHEORIE} }

\author{  {VON}\\ {\bf \huge KORN\'{E}L LAEWY (L\'{A}NCZOS)}\\ 
       {\normalsize ASSISTENT AN DER TECHN. HOCHSCHULE}\\ ~\\ ~\\ ~\\
                    {\large BUDAPEST, 1919.}\\
             VERLAGSBUCHHANDLUNG JOSEF N\'{E}METH\\
        {\footnotesize I., FEH\'{E}RV\'{A}RI-\'{U}T 15.} }

\date{~\\ Capture and typesetting by\\
          {\bf Jean-Pierre Hurni}\\
                    ~\\
            Abstract and preface by\\
            {\bf Andre Gsponer}\\
           ~~\\
ISRI-04-06.12 ~~~ \today}

\maketitle

\newpage

~~\\

\begin{center}
{\bf  \vspace{-2.0cm}{\huge  THE\\
                   FUNCTIONAL THEORETICAL\\
                   RELATIONSHIPS OF THE\\
                       HOMOGENEOUS\footnote{In contemporary language, the
word `Aether' used by Lanczos in the title of his dissertation should be
translated by `vacuum'.  However, in view of the ideas developed by Lanczos in his thesis, a possibly better translation should be `homogeneous.' An alternate translation of the full title could be: `The relations of the homogeneous Maxwell's equations to the theory of functions.'}\\
                    MAXWELL EQUATIONS\\}
                           ~\\
                           ~\\
     {\large A CONTRIBUTION TO THE THEORY OF \\
                RELATIVITY AND ELECTRONS\\
                           ~\\
       ``Dedicated to Albert Einstein and Max Planck,\\
        the two great standard-bearers of constructive speculation''}~\footnote{``Den hohen Fahnentr\"agern der Konstruktive Spekulation, Albert Einstein und Max Planck, in ehrerbietigster Hochachtung gewidmet.'' Letter of Lanczos to Einstein, 3 December 1919, {\bf in} W.R.\ Davis \emph{et al.}, eds., Cornelius Lanczos Collected Published Papers With Commentaries (North Carolina State University, Raleigh, 1998) Volume {\bf I}, page 2-40.} }\\
                           ~\\
                           ~\\
                      ~\\  {BY}\\
                           ~\\
           {\bf \huge CORNELIUS LANCZOS\\ 
   {\footnotesize \vspace{-0.4cm} ASSISTANT AT THE INSTITUTE OF TECHNOLOGY}\\
                           ~\\
                   {\large BUDAPEST\\
                          1919}\\}
 \end{center}

\newpage 

\vspace{0cm}

\begin{center} {\Huge {\bf Abstract}} \end{center}

~\\

\begin{quote}
{\large The thesis developed by Cornelius Lanczos in his doctoral dissertation is that electrodynamics is a pure field theory which is hyperanalytic over the algebra of biquaternions.  In this theory Maxwell's homogeneous equations correspond to a generalization of the Cauchy-Riemann regularity conditions to four complex variables, and electrons to singularities in the Maxwell field.  Since there are no material particles in Lanczos electrodynamics, the same action principle applies to both regular and singular Maxwell fields.  Therefore, the usual action integral of classical electrodynamics is \emph{not} an input in that theory, but rather a consequence which \emph{derives} from the application of Hamilton's principle to a superposition of two or more homogeneous Maxwell fields.  This leads to a fully consistent electrodynamics which, moreover, can be shown to be finite. As byproducts to this remarkable thesis Lanczos anticipated the Moisil-Fueter theory of quaternion-analytic functions by more than ten years; showed that Maxwell's equations are invariant in both spin-1 and spin-1/2 Lorentz transformations; that displacing a singularity into imaginary space adds an intrinsic magnetic-like field to its electric field; and that his theory does even include gravitation --- although not in the general relativistic form of Einstein to whom Lanczos dedicated his dissertation.}
\end{quote}

\newpage

~\\
\vspace{2\baselineskip}

\noindent{\Huge  {\bf  Preface}}

~\\

\noindent{Lanczos}'s monumental doctoral dissertation is now at last available in typeseted form thanks to Dr.\ Jean-Pierre Hurni who took on himself the painstaking task of keying in Lanczos's manuscript, as well as of resolving the many problems which arise when capturing a text handwritten in German by a Hungarian in 1919.

A facsimile of Lanczos's handwritten dissertation is included in the Appendix of the \emph{Cornelius Lanczos Collected Published Papers With Commentaries} \cite{DAVIS1998-}.  It is therefore advisable that readers finding a problem with the present typeseted version have a look at the manuscript, and possibly let us know of any mistake which should be corrected in a revised version of this transcription.\footnote{In the traditional spirit of typography, Jean-Pierre Hurni and myself have made a few formal alterations to Lanczos's manuscript in order to improve the readability of its typeseted version.  This is why we have modified or added some punctuation, put in full words some abbreviations, numbered the equations and figures, replaced ``$0$'' and ``$1$'' by ``Null'' and ``Eins,'' introduced the modern notation $\A^*$ for complex conjugation, etc. However, we have not interfered with a few peculiarities (such as Lanczos's generous use of colons, or minimal use of punctuation in formulas) in order not to go beyond what is permissible from a strict typographical point of view.}

In the proceedings of the 1993 \emph{Cornelius Lanczos International Centenary Conference}, George Marx, President of the E\"otv\"os Physical Society of Hungary, gives a number of background details on Lanczos's dissertation, including excerpts of the correspondence between Lanczos and Einstein related to it \cite{MARX-1993-}. In the \emph{Lanczos Collection} there is also a commentary by myself and Jean-Pierre Hurni on that dissertation \cite{GSPON1998A}.  While this commentary was written in 1994, a more elaborate version of it is now available in electronic form \cite{GSPON2004E}.  This expanded commentary is showing, in particular, that Lanczos electrodynamics leads to a fully consistent and finite electrodynamic theory, in which the \emph{finite} mass appearing in the usual action integral of electrodynamics, and in the Abraham-Lorentz-Dirac equation of motion, is neither the ``mechanical'' nor the ``electromagnetic'' mass, but strictly the \emph{inertial} mass of D'Alambert and Einstein.  Therefore, ``Lanczos's electrodynamics'' is not just an alternate formulation of classical electrodynamics, but a full fledged field theory which encompasses classical electrodynamics as well as some fundamental aspects of general relativity and quantum theory.

Since substantial attention is given in these commentaries to the contemporary relevance of Lanczos's functional theory of electrodynamics, it will be enough as an introduction to Lanczos's dissertation to highlight, chapter by chapter, the main conclusions reached by him in the development of his thesis:

\begin{enumerate}

\item In Chapters 1 to 3 Lanczos shows how quaternions are ``exceptionally well adapted to the study of the [four-dimensional universe and] general nature of an arbitrary Lorentz transformation'' \cite[p.304]{LANCZ1949-}, a demonstration he will repeat in the first of his 1929 papers on Dirac's equation \cite{LANCZ1929A}, and in chapter IX of his 1949 book on the variational principles of mechanics \cite{LANCZ1949-}.  To this end he devotes Chapter 1 to the introduction of the two basic types of monomials, that he calls \emph{vector} and \emph{versor}, which arise in the covariant formulation of four-dimensional objects with quaternions.\footnote{This classification does not take spinors into account, even though Lanczos will come very close to their discovery in Chapter 2 and 4.}  This enables him to introduce the definition of the quaternion product, equation (1.3), in a very elegant and natural way, which leads him to define the versor $A\CON{B}$ as the product of a vector $A$ by the quaternion conjugate of a vector $B$.  Lanczos's vector is therefore what is commonly called a ``four-vector'' (e.g., the four-dimensional position $it+\vec{x}$, or the energy-momentum quaternion $\mathcal{E}+i\vec{p}$\,) which in the general case have four complex components transforming as contra- or co-variant vectors is tensor calculus.  On the other hand, Lanczos's versor is a quaternion whose vector part is what is commonly called a ``six-vector,'' (e.g., the complex combination $\vec{E}+i\vec{H}$ of the electric and magnetic field vectors) which in the general case have six real tensor components transforming as an antisymmetric four-tensor of rank two, and whose scalar part is an invariant complex number.  Lanczos's ``versor'' is therefore a slight generalization of Hamilton's original concept of versor, a generalization which stresses the fact that by multiplying four-vectors the resulting monomials $A\CON{B}C\CON{D}...$ transform covariantly either as vectors or as versors.\footnote{Since the vector part of a versor is a six-vector, and its scalar part an invariant, Lanczos's notion of a versor is not very useful in practice because (using contemporary field theory language) the former corresponds to a ``vector field,'' and the later to a ``scalar field,'' which should be segregated rather than united according to field theory.  For this reason we recommend to avoid the use of the terms versor and vector in Lanczos's sense, but to use instead the terms ``six-vector,'' ``four-vector,'' and ``invariant'' with their usual meaning to specify how these objects behave in a Lorentz transformation; as well as to use the unqualified term ``vector'' in Hamilton's original sense, that is for a real three-dimensional object $\vec{v}$.}

\item In Chapter 2 Lanczos shows --- equations (2.12) and (2.23) --- that a general four-dimensional rotation (which combines a spatial rotation and a Lorentz boost) can be written $p \A q$ with $q=\CON{p}^*$ for a ``vector'' (i.e., four-vector) and  $p \A \CON{p}$ for a ``versor'' (i.e., six-vector).  In order to obtain these expressions Lanczos starts right away by showing that the left- and respectively right-multiplications by a biquaternion (operations that Lanczos calls $\mathcal{P}$- and respectively $\mathcal{Q}$-transformations) are directly related to orthogonal transformations.  He therefore recalls the remarkable property of quaternions (already discovered by Hamilton) which is to provide an explicit spinor decomposition of the general four-dimensional orthogonal transformation, so that each $\mathcal{P}$- or $\mathcal{Q}$-transformations taken by themselves are noting but spinor rather than tensor transformations.\footnote{Of course, in 1919, Lanczos was most certainly not aware of this interpretation, which stems from the discovery of double-valued representations of the rotation group, sometimes after their classification by Elie Cartan in 1913.}

\item  Chapter 3 is a superbe generalization of the fundamental axioms and theorems of complex function analysis to quaternions. In a very lucid and concise manner Lanczos does what will be rediscovered by Moisil and Fueter in 1931.  In particular, the \emph{Cauchy-Riemann-Lanczos-Fueter regularity conditions} correspond to equation (3.5) or (3.6), and the \emph{Cauchy-Lanczos-Fueter integration formula} to equation (3.19), see references in \cite{GSPON1998A}.

\item While Chapter 1 and 2 introduced quaternions as a means to endow space-time with the powerful algebraic structure provided by Hamilton's quaternion algebra, and Chapter 3 introduced quaternion analyticity as a first step towards a biquaternion\footnote{In quaternion terminology the prefix `bi' means that a quantity is complexified.} theory of analytical fields over space-time, the first major step in Chapter 4 is Lanczos's recognition that the identification
$$
                             t = i\tau ~~,  \eqno(1)
$$
which leads from the \emph{Cauchy-Riemann-Lanczos-Fueter regularity conditions} (3.6) to the \emph{homogeneous Maxwell's equations} (4.7), is most important for the understanding of the physical nature of space-time.  Indeed, it is through this identification, i.e., the definition of time as an intrinsically \emph{imaginary} quantity, that null-quaternions\footnote{That is non-zero four-dimensional objects whose norm is zero, something that is only possible in complexified four-space.} --- and thus four different types of spinors --- enter into the description of space-time objects.  Ultimately, as will later be explained by Lanczos, the origin of ``$i$'' in quantum physics stems from this identification, something that he will repeat until the end of his professional career: ``For reasons connected with the imaginary value of the fourth Minkowskian coordinate $ict$, the wave mechanical functions assume \emph{complex values}'' \cite[p.268]{LANCZ1970-}.\footnote{In his later years, Lanczos will insist that the Minkowskian metric should not enter theory simply as an empirical fact, but rather be deduced from a more fundamental theory based on a positive-definite four-dimensional Riemannian metric. This lead him to investigate the structure of such a theory, and to find out --- in particular --- an explanation for Einstein's photon hypothesis of 1905, see \cite{LANCZ1963A}, and even to derive the entire Maxwell-Lorentz type of electrodynamics \cite{LANCZ1974B}.}

 Then, having shown the direct correspondence between Maxwell's homogeneous equations and biquaternion analytic function over complexified space-time, Lanczos takes note that such hyperanalytic functions are not restricted to just vector functions such as Maxwell's bivector field $\vec{E}+i\vec{H}$, but that they may be any versor field $\mathcal{F}$ composed of a scalar and a vector.  Moreover, Lanczos realizes that Maxwell's homogeneous equations do not specify by themselves the full behaviour of such a field in a Lorentz transformation: Only the  $\mathcal{P}$-transformation is implied by them, while --- see his equation (4.9) --- the $\mathcal{Q}$-transformation operator $\A q$ may be multiplied from the right by any quaternion $q_0$ with unit norm, i.e., $|q_0| = 1$.

Lanczos therefore very consciously realized that the homogeneous field equations (4.7) are invariant, besides the Lorentz group, under transformations which correspond, in modern language, to the three parameter group $SU(2) \sim \mathbb{H}/ \mathbb{R}$.\footnote{``Eine dreidimensionale Mannigfaltigkeit.'' This is quite remarkable, and illustrative of the insight provided by the quaternion formalism: $U(1)$ phase transformations will not explicitly be considered before G.Y.\ Rainich in 1925, and interpreted as gauge transformations by H.\ Weyl in 1929, while $SU(2)$ non-abelian phase transformations will not be discussed before C.N.\ Yang and F.\ Mills in 1954.}  He therefore concludes that $\mathcal{F}$ may correspond  to either a six-vector (i.e., $qq_0=\CON{p}$), in which case the field is just the electromagnetic field, or to a four-vector (i.e., $qq_0=\CON{p}^*$, see end of Chapter 2), in which case Lanczos proposes that the field could correspond to the gradient of a scalar potential, which he associates to gravitation.

Unfortunately, Lanczos did not consider the case $qq_0=1$ which corresponds to the trivial identity $\mathcal{Q}$-transformation:  This would have led him to contemplate the possibility of massless spin-1/2 particles!  Nevertheless, right after the discovery of Dirac's equation in 1928, Lanczos will remember that, and take advantage of the quaternion formalism to fully explain the space-time covariance properties of spin-1 and spin-1/2 wave equations \cite{GSPON1994-,GSPON1998B}, months before van der Waerden and others will do the same, albeit only implicitly, by introducing `dotted' and `undoted' indices into the tensor formalism.



\item Chapter 5 is very brief and most important:  If electrons are to be interpreted as moving singularities of the Maxwell field, and if these singularities are to be defined by the vacuum (i.e., homogeneous) Maxwell's equations, then there is no room for the inhomogeneous Maxwell's equations in such a theory.  As stated twice by Lanczos, the homogeneous Maxwell's equations should not be linked to right-hand members which are ``foreign''\footnote{``fremd''} to the function, and ``in contrast to the true mathematical spirit of these equations.''\footnote{``in einem merkwürdigen Kontrast zu dem wirklichen mathematischen Geiste dieser Gleichungen.''}  Therefore, in contradistinction to the paradigm which is still prevalent today, there are \emph{no} currents, \emph{no} sources, in Lanczos's electrodynamics!  Summarizing his strictly logical interpretation of what a pure field theory is, Lanczos states: \emph{``Matter represents the singular points of the  corresponding functions  which  are  determined  by  the  vacuum  differential equations.''}\footnote{\emph{``Die Materie repr\"asentiert die singul\"aren Stellen derjenigen Funktionen, welche durch die im Aether gültigen Differentialgleichungen bestimmt werden.''} Underlined by Lanczos.}  Then, pushing his reasoning to its logical end, Lanczos explains that his theory resolves the fundamental paradox of the ``\,Theory of Electrons''\footnote{The expression ``\,Theory of Electrons'' which appears in the subtitle of Lanczos's dissertation refers to the theory of Lorentz, and others, in which electrons are postulated to be material particles of finite or vanishing radius.}  --- ``How can an object made out of strongly repulsive forces hold together''\footnote{``wie ein Gebilde bei lauter abstossenden Kr\"aften zusammengehalten werden kann''}~?~--- simply because there is no problem of stability in a field theory where an electron is just a singularity.

\item In Chapter 6 Lanczos starts by considering the fundamental particular solution to the potential equation, the Li\'enard-Wiechert potential of an electron in relativistc motion, and by remarking that it is possible to derive a whole series of new particular solutions by simply derivating them with respect to the coordinates.  He therefore concludes that ``An electron can be seen as a structure with an infinite number of degrees of freedom,''\footnote{``Ein Elektron kann somit als ein Gebilde mit unendlich vielen Freiheitsgraden angesehen werden.''} which means that his theory can be applied to atomic, molecular, as well as macroscopic structures.

However, now that the stage is set, an action principle is required to define the dynamics.  To this end,  Lanczos soon discovers that the only covariant way to apply Hamilton's principle is to write the action as 
$$
   \REA \iiiint dx\,dy\,dz\,d\tau~ \mathcal{F}\CON{\mathcal{F}} ~~, \eqno(2)
$$
where $\mathcal{F}$ is the total electromagnetic field of all electrons and external fields, and $\mathcal{F}\CON{\mathcal{F}}$ the scalar product of this total field by itself.  For example, writing the self-field of some electron as $\mathcal{F}_i$, the action corresponding to its interaction with a given external field $\mathcal{F}_e$ will be
$$
 \REA \iiiint  dx\,dy\,dz\,d\tau~ 
         (     \mathcal{F}_i  +      \mathcal{F}_e )
         (\CON{\mathcal{F}_i} + \CON{\mathcal{F}_e}) \eqno(3)
$$
which trivially leads to the expression\footnote{The operator ``$\Scal\Q$'' means that we take the scalar part of the bracketed quaternion expression.} 
$$
 \REA \iiiint dx\,dy\,dz\,d\tau~
            \Scal\Bigl[\mathcal{F}_i \CON{\mathcal{F}_i}
                   + 2 \mathcal{F}_i \CON{\mathcal{F}_e}
                   +   \mathcal{F}_e \CON{\mathcal{F}_e}
                 \Bigr]  ~~.                     \eqno(4)
$$
Therefore, if Lanczos's thesis is correct, and provided all integrals are feasible and finite, one should be able to \emph{derive} the standard classical electrodynamics action integral, which in that case should simply be 
$$
     m   \int    d\tau~
  +  e   \int    d\tau~  \Scal\Bigl[\mathcal{U}_i \CON{\Phi_e}\Bigr]
  + \REA \iiiint dx\,dy\,dz\,d\tau~
                   \mathcal{F}_e \CON{\mathcal{F}_e}              \eqno(5)
$$
where $m$, $e$, and $\mathcal{U}_i$ are the mass, charge, and four-velocity of the electron, and $\Phi_e$ the four-potential of the external field  $\mathcal{F}_e$.  In practice, if the integrations in equation (4) are made in the ``standard way,'' that is as volume integrals using for $\mathcal{F}_i$ the electromagnetic field derived from the Li\'enard-Wiechert potentials of an arbitrarily moving electron, one immediately finds out that in the general case the first two terms diverge because of the singularity at the origin of the field.  The reason is that the ``standard way'' does not take the full nature of electromagnetic singularities into account, a point that Lanczos acutely understood:  The four-dimensional integrations should be made in the spirit of field theory, that is as surface integrals, something that is always possible since the homogeneous Maxwell's equations enable to use Gauss's theorem --- equation (6.14) --- to transform the volume integrals into surface integrals.\footnote{This is precisely what is done in order to derive the Cauchy-Lanczos-Fueter integration formula (3.19) from the Cauchy-Riemann-Lanczos-Fueter analyticity conditions (3.6).} Thus, instead of equation (3), Lanczos is led to consider the integral 
$$
 \REA \iiint \Scal\Bigl[ (\Phi_i  +    \Phi_e) d^3\Sigma
             (\CON{\mathcal{F}_i} + \CON{\mathcal{F}_e}) \Bigl] 
                                                                   ~~, \eqno(6)
$$
i.e., his equation (6.16), where $d^3\Sigma$ is now a closed hypersurface to be carefully chosen in accord with the locations of the singularities, and possibly with other boundary conditions.  Unfortunately, while I was able to show with Jean-Pierre Hurni that equation (6) does indeed lead to equation (5) --- see our commentary \cite{GSPON2004E} --- Lanczos was not able (or possibly did not even attempt) to perform the required integrations using a closed hypersurface and to show that the result is finite.

Nevertheless, Lanczos properly grasped all the main ideas, and was only stopped by the purely technical difficulty of calculating non-trivial three-surface integrals.  In particular, Lanczos fully realized that the proper choice of the hypersurface bounding the domain of integration was a very important question, since it is precisely the values of the field on this boundary which determine the value of the function within that domain --- equation (3.19).  In this respect, it is worth stressing that this crucial point was essentially forgotten in the twenty years that followed Lanczos dissertation, until Paul Weiss \cite{WEISS1936-} rediscovered the importance of general surfaces in the calculation of four-dimensional quantum action integrals, a point that opened the way to the later theories of Tomonoga, Schwinger, \emph{et al.}, which led to modern quantum electrodynamics --- see references in \cite{GSPON1998A}.\footnote{In the same vein Paul Weiss developed powerful methods for the explicit calculation of four-dimensional surface integrals, using for this purpose the biquaternion algebra to make explicitly the spinor decomposition of four-vectors and six-vectors \cite{WEISS1941-}.}

In fact, in the course of this chapter, after writing down the four-dimensional form of the action integral --- equation (6.7) --- and after applying Gauss's theorem --- equation (6.14) --- Lanczos discusses the boundaries to be considered in great details.  He even summarizes his intuitive understanding of the cosmological implications of his field theory in a full page figure, Fig.~6.1, which emphasizes the importance of keeping all integrations within the bounds of the past and future light-cones.  It is therefore unfortunate that the last paragraph of Chapter 6 is an act of resignation, in which he accepts --- without any mathematical justification --- the prevalent dogma that the self-energy integral should be divergent. 

\item In Chapter 7, having accepted in the previous chapter the apparently unavoidable divergent nature of the self-energy of a point electron, Lanczos has a stroke of genius: What about displacing the position of the electron off the world-line into complexified space? In that case a purely electric field in the rest-frame gets an additional magnetic field contribution, and a simple calculation shows that the self-energy integral is finite, e.g., zero in the case of an imaginary translation --- a model that Lanczos calls the \emph{circle electron}, which will be later rediscovered by others \cite{NEWMA1973-}.  Lanczos therefore concludes that the electron's self-energy contribution to the action integral is not necessarily infinite, a possibility he will take for granted in the following chapter.  At this point two comments are in order: 

First. If translating the position of an electron into imaginary space does indeed add an imaginary component to the electric field, this imaginary component is in fact not a magnetic field, but rather a \emph{mesomagnetic} field which in a space-reversal transforms as a polar rather than axial vector \cite{GSPON2004C}.  However, at Lanczos's time the problems associated with the physical interpretation of improper Lorentz transformations such as space-reversal were far from being fully understood. (This had to wait until the late 1920s early 1930s, if not the discovery of parity violation in the mid 1950s.)  Lanczos should therefore not be blamed for that.

Second.  The particle spectrum in Lanczos's electrodynamics is potentially very large, and possibly sufficiently rich to include all known elementary particles.  This is due to the possibility of shifting the position of the singularity into complexified space; to the previously noted feature that the Cauchy-Riemann-Lanczos-Fueter conditions allow for singularities and fields other than just electromagnetic, e.g., six-vector, four-vector, or spinor; to the infinite dimensional character of the singularities themselves (see beginning of Chapter 6); to the possibility that singularities might be clusters of more elementary singularities; etc.

\item At the beginning of Chapter 8 Lanczos stresses once again the importance of boundary surfaces in the calculation of the action integral: ``\,The contribution of these surfaces can in no way simply be ignored, even if the boundaries lay at infinity.  It is much more probable that the boundaries have a characteristic role to play. [...] If the field-theoretical point of view is correct, the boundaries must also have a field-theoretical meaning.''\footnote{``Der Beitrag dieser Fl\"ache darf keineswegs einfach Null gesetzt werden, selbst wenn die Grenzen im Unendlichen liegen. Es ist vielmehr wahrscheinlich, dass der Grenzfl\"ache eine charakteristische Rolle zukommen wird. Wir haben wohl die Grenzkegelfl\"achen des m\"oglichen Raumes relativtheoretisch gerechtfertigt, wenn aber der funktionentheoretische Standpunkt richtig ist, so müssen diese Grenzen auch funktionentheoretische Bedeutung haben.''} Having said this, Lanczos briefly speculates on the possible cosmological implications of light-cone related singularities,\footnote{His concept, summarized in the full page figure, Fig.~6.1, of all world-lines in the Universe stemming from a single original singularity, and focusing on a single final singularity, the ``big bang'' and the ``big crunch,'' the ``alpha'' and the ``omega,'' is an omnipresent idea in the Judeo-Christian culture.} and then only, almost reluctantly, goes to the main topic of the chapter: The derivation of the equations of motions of an electron in a gravitational or an electromagnetic field. To this end he assumes that the self-interaction term in the action (which he calls the ``electron's Hamiltonian function'') is zero (or at least finite and negligible) in order to focus on the interaction term.  After some lengthy calculations he succeeds in deriving Minkowski's generalization of the Newton force --- equation (8.22) --- as well as the Lorentz's force  --- equation (8.32).  However, in sharp contrast with the rest of his dissertation, the calculations are botched up, as if Lanczos had been in a haste, or had little interest in going through an ``applied'' rather than ``theoretical'' development...  Thus, while there is little more than a confirmation of two anticipated results in it, that chapter tells us a lot about Lanczos's psychology and preference to think about ``fundamental'' rather than ``utilitarian'' questions.

\item In the conclusion Lanczos first summarizes his hope: that, as a result of some variation,  a good theory should not only predict the electric charge and the mechanical mass of an electron, but also the relations between them; and his regret: that, in this respect, his own theory does not go significantly beyond the ``\,Theory of Electrons.''  He therefore goes on to his conclusion, in the form of a very lucid and personal assessment of his dissertation which is worth quoting \emph{in extenso}:

\begin{quote}
``\,The theory which is here sketched is meant to be a contribution
to the constructive formulation of modern physical theory, in
the sense that has particularly being introduced by the works of
Einstein. Its value, or lack thereof, should therefore not be
judged according to practical positivist-economic principles ---
because it does not pretend to provide any simple `working
hypothesis.'  Its convincing power --- when I am not missled by
my  subjectivity --- does not lie in `striking proofs,' but in
the consistency and non-arbitrariness of its construction, by
which, in capturing the proper soul of Maxwell's equations, the
theory of Maxwell fused with the theory of relativity, it leads
to electrons in a natural way.  This systematic simplicity and
necessity provides the basis for my view of its superiority over
the usual theory of the electron.  I have not gone here into the
details, but just into the outlines.  More precisely, I have been
concerned with merely preparing a direct way,  the path of
which when followed may possibly open new perspectives into the
inscrutable depths of Nature.'' 
\end{quote}

\item Finally, in a brief \emph{postscript}, Lanczos reports on his afterthought that, in actual facts, the introduction (in Chapter 6) of Hamilton's principle as a separate axiom of his theory was not necessary.  Indeed, it turns out that the variation of the action for Maxwell's field (which in biquaternionic functional theory is in direct correspondence with a similar variation principle) necessarily leads to the Cauchy-Riemann-Lanczos-Fueter regularity conditions, so that by taking these conditions\footnote{Together, as stressed by Lanczos, with the boundary conditions.} as ``fundamental equations''\footnote{``Grundgleichungen''}  one implicitly includes Hamilton's principle, and vice versa.\footnote{For a related correspondence between Hamilton's principle, classical mechanics in Hamiltonian form, and a biquaternionic regularity condition see \cite[Section~5]{GSPON1993-}.}  This proves the internal consistence of Lanczos's electrodynamics, and demonstrates that the scope of Lanczos's theory goes beyond standard electrodynamics and mechanics. 

\end{enumerate}

A remakable formal aspect of this dissertation (as well as of Lanczos's later papers using quaternions) is its style:  It is definitely \emph{modern} in the sense that throughout his dissertation Lanczos deals with complex scalars, vectors, and quaternions (i.e., 2 to 8 dimensional objects over the reals) as \emph{whole symbols} --- which he mixes freely --- without using the antiquated quaternion notations, definitions, and vocabulary derived from Hamilton's original papers.  While this makes Lanczos's dissertation and other quaternion papers more readable and accessible to us, it must have made them look foreign to the traditional quaternion users at Lanczos's time, so that few of them took the trouble of reading his papers.

To conclude this preface, let us recall that Lanczos dedicated his dissertation to Einstein --- who accepted the dedication --- and that this was the beginning of a life-long correspondence and occasional collaboration between them.  In particular, when Lanczos would become Einstein's personal assistant in 1928, he will return to quaternions in order to show how the relativistic spin-1/2 wave equation recently found by Dirac could in fact be \emph{derived} from a quaternion field theory which implied that elementary particles such as electrons should have both spin \emph{and} isospin, so that Dirac's equation on its own would concern only half of the  elementary particles spectrum.  Again, just like with his doctoral dissertation, nobody will really try to understand Lanczos's prodigious logical deductions.\footnote{Except possibly Einstein who, with his next assistant Walter Mayer, would produce his own version of quaternions (the now forgotten ``semivectors'' \cite[p.112]{MURNA1945-}) and further generalize the equation from which Lanczos derived Dirac's equation, in order to lift its mass-degeneracy \cite{GSPON1994-,GSPON1998B} so that \emph{massless} spin-1/2 particles would have to exist on the same footing as electrons --- an idea strongly rejected at the time by Wolfgang Pauli and others.}

~\\

\indent \hspace{5.8cm}       Andre Gsponer\\
\indent \hspace{5.8cm} \emph{Associate editor of the Lanczos Collection}

\newpage

\noindent{\Huge {\bf Acknowledgments}}

~\\

\noindent We are greatly indebted to Prof.\ William R.\ Davis for having taken the initiative of organizing the 1993 \emph{Cornelius Lanczos International Centenary Conference}, which gave Andre Gsponer the opportunity to make a photocopy of Lanczos's dissertation, to talk to Prof.\ George Marx about Lanczos's dissertation and related quaternion work, and to be invited at lunch by Prof.\ John Archibald Wheeler --- who apparently was the only person to come in at the minisymposium at which Andre Gsponer gave his talk especially to listen to it \cite{GSPON1994-} --- in order to discuss Lanczos's ideas on classical electrodynamics and Dirac's equation.  We also thank Mr.\ Jean-Claude Ziswiler, Dr.\ J\"org Wenninger, and Prof.\ Gerhard Wanner for their help in finding the correct transcription of some badly readable parts of the dissertation;  as well as Dr.\ Jacques Falquet for scanning and computer processing the hand-drawn illustrations of Lanczos's dissertation.

\selectlanguage{german}

\newpage\thispagestyle{empty}
\begin{figure}
\begin{center}
\resizebox{16cm}{!}{ \includegraphics{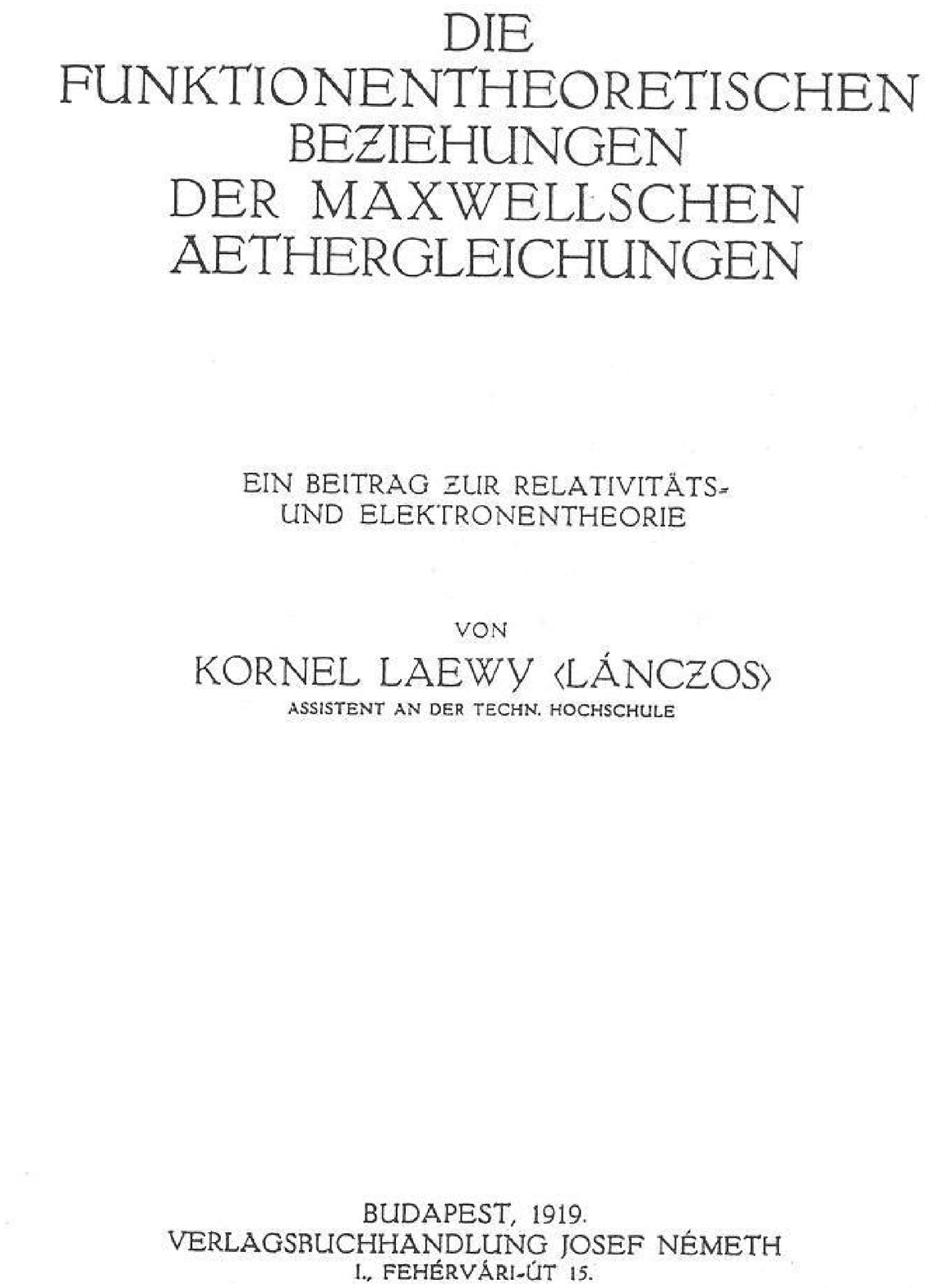}}
\end{center}
\caption{Titel}
\end{figure}

\tableofcontents

\medskip
\medskip
\noindent{\bf ~ ~ ~ ~ Nachtrag. \hspace{10.6cm}  56}

\listoffigures

\chapter{Spezielle Transformations\-eigenschaften
                des vierdimensionalen Raumes.
                Beziehungen zu den Quaternionen.}

   Die M{\scriptsize INKOWSKI}sche Vektoranalysis beruht auf den linearen geometrischen Gebild\-en des E{\scriptsize UKLID}ischen Raumes. In ihr spielt die Zahl der Dimension keinerlei bevorzugte Rolle, das Raum-Zeit-Kontinuum mit seinen vier Dimensionen bildet einen speziellen Fall des E{\scriptsize UKLID}ischen Raumes mit der allgemeinen Dimension: $n$.  Es besitzt aber gerade der vierdimensionale Raum in Hinsicht auf die orthogonalen Transformationen (Drehung) Eigenschaften, welche ihn allen anderen R\"aumen gegenüber auszeichnen. Diese Eigenschaften, die mit den H{\scriptsize AMILTON}schen Quaternionen nahe zusammenh\"angen, erlauben einerseits eine grunds\"atzliche Vereingleichung und einheitlichen Aufbau auf die Quaternionenrechnung für die gesamte vierdimensionale Vektoranalysis, andererseits erm\"oglichen sie durch die Anpassung an den speziellen Charakter der Dimension $n=4$ in Anwendung auf das elektromagnetische Feld die Grenzen der Feldtheorie in naturgem\"asser Weise über den M{\scriptsize INKOWSKI}schen Rahmen hinaus zu erweiten.

   Das scalare ebenso wie das vektorielle Produkt zweier Vektoren l\"asst sich durch die Forderung einführen, aus den Komponenten zweier Vektoren ein System von quadratischen Gebilden zu konstruieren, mit der Eigenschaft, dass bei einer Drehung des Achsenkreuzes das neue System mit dem alten in homogen linearer Weise zusammenh\"ange. Es bildet das skalare Produkt durch seine Invarianz ein solches System, das vektorielle Produkt mit $\binom{n}{2}$ Gliedern ein anderes. Damit sind die M\"oglichkeiten im allgemeinen ersch\"opft (abgesehen von dem allgemeinsten, aber trivialen Fall, dass auch alle überhaupt m\"oglichen Produkte zweier beliebiger Komponenten zusammengenommen ebenfalls ein verlangtes System ergeben). Gerade bei der Dimension $n=4$ ist aber noch ein anderes System --- und zwar ein dreigliedriges --- konstruierbar.

   Wir schreiben das vektorielle Produkt von den zwei Vierervektoren: $(X_1,Y_1,$ $Z_1,T_1)$ und $(X_2,Y_2,Z_2,T_2)$ mit den üblichen Bezeichnungen hin, ebenso auch den ``dualen'' Vektor. Beide sind Sechservektoren:
\begin{gather}
\left.
\begin{array}{c}
\mathcal{F}_{yz} = Y_1 Z_2 - Y_2 Z_1  \\
\mathcal{F}_{zx} = Z_1 X_2 - Z_2 X_1  \\
\mathcal{F}_{xy} = X_1 Y_2 - X_2 Y_1  \\
\mathcal{F}_{xt} = X_1 T_2 - X_2 T_1  \\
\mathcal{F}_{yt} = Y_1 T_2 - Y_2 T_1  \\
\mathcal{F}_{zt} = Z_1 T_2 - Z_2 T_1
\end{array}
\quad \right\}
\hskip 0.8cm
\left.
\begin{array}{c}
\mathcal{F}_{yz}^\sharp = \mathcal{F}_{xt} = X_1 T_2 - X_2 T_1 \\
\mathcal{F}_{zx}^\sharp = \mathcal{F}_{yt} = Y_1 T_2 - Y_2 T_1 \\
\mathcal{F}_{xy}^\sharp = \mathcal{F}_{zt} = Z_1 T_2 - Z_2 T_1 \\
\mathcal{F}_{xt}^\sharp = \mathcal{F}_{yz} = Y_1 Z_2 - Y_2 Z_1 \\
\mathcal{F}_{yt}^\sharp = \mathcal{F}_{zx} = Z_1 X_2 - Z_2 X_1 \\
\mathcal{F}_{zt}^\sharp = \mathcal{F}_{xy} = X_1 Y_2 - X_2 Y_1
\end{array}
\quad \right\}
\end{gather}
Da der duale Vektor mit dem ursprünglichen kovariant ist, so gilt dies auch von der Summe oder der Differenz beider. Dabei kommen nun in beiden F\"allen nur drei voneinander verschiedene Gr\"ossen vor, n\"amlich:
\begin{equation}
\left.
\begin{array}{c}
\mathcal{F}_{xt} \pm \mathcal{F}_{yz} \\
\mathcal{F}_{yt} \pm \mathcal{F}_{zx} \\
\mathcal{F}_{zt} \pm \mathcal{F}_{xy}
\end{array}
\quad \right\}
\end{equation}
(Die zwei Vorzeichen sind als entweder-oder zu verstehen.) Dieses System von drei Gliedern hat also ebenfalls die Eigenschaft, mit dem entsprechenden im transformierten System homogen und linear zusammenzuh\"angen. Bedenken wir nun, dass der Raumteil eines Vierervektors, entweder reell oder rein imagin\"ar ist, entsprechend der Zeitteil umgekehrt imagin\"ar, bzw., reell, so sehen wir, dass die zuletzt-hingeschriebenen drei Gr\"ossen komplexe Zahlen vorstellen. Eine komplexe Zahl charakterisiert aber sowohl ihren reellen, wie ihren imagin\"aren Teil, so dass --- bei Zulassung komplexer Zahlen --- der ursprüngliche Sechservektor vollkommen durch diese drei Gr\"ossen ersetzbar ist. Nehmen wir als vierte das skalare Produkt der beiden Vektoren hinzu, so erhalten wir das folgende, alle zwei Multiplikationsarten enthaltende System:
\begin{align}
\left.
\begin{array}{c}
X_1 T_2 - X_2 T_1   \pm (Y_1 Z_2 - Y_2 Z_1) \\
Y_1 T_2 - Y_2 T_1   \pm (Z_1 X_2 - Z_2 X_1) \\
Z_1 T_2 - Z_2 T_1   \pm (X_1 Y_2 - X_2 Y_1) \\
X_1 X_2 - Y_1 Y_2    + ~ Z_1 Z_2 + T_1 T_2 ~~~
\end{array}
\quad \right\}
\end{align}
Betrachten wir nun den Vektor: $(X_1,Y_1,Z_1,T_1)$ als Quaternion, --- wobei der sogenannte ``skalare Teil'' durch den Zeitteil des Vektors representiert wird --- und multiplizieren wir mit die Quaternion: $(-X_2,-Y_2,-Z_2,T_2)$, so erhalten wir als Produkt eine Quaternion, deren Komponenten eben durch die hingeschriebenen Werte auch in Hinsicht der Reihenfolge dargestellt werden, und zwar, wenn wir das untere (negative) Zeichen w\"ahlen. Mit dem positiven Zeichen hingegen, wenn die Reihenfolge derselben zwei Quaternionen als Faktoren die entgegengesetzte ist. Dieses Produkt kann jedoch nicht mehr einfach als Vektor bezeichnet werden, weil es ja bei einer orthogonalen Transformation mit den Vektorkomponenten nicht kovariant ist. Andererseits h\"angen aber doch die alten und neuen Komponenten des Produktes in homogen linearer Weise von einander ab, --- und das ist ja in Hinsicht auf das Relativit\"atsprinzip das ausschlaggebende --- nur ist die Matrix der Transformation von der ursprünglichen Matrix verschieden. Wir gelangen so zu einer Erweiterung des ursprünglichen Vektorbegriffes, welche eine einheitliche Zusammenfassung der Vierer- und Sechservektoren, sowie auch der Skalaren erlaubt. Wir setzen fest, dass die Zahl der Komponenten durchwegs vier sei und diese Komponenten sollen sich bei einer beliebigen orthogonalen Transformation in homogen linearer Weise transformieren, wobei die Koeffizienten der Transformation von dessen der Koordinatentransformation verschieden sein dürfen. Es sei mir erlaubt der Kürze halber einen solchen Inbegriff von vier Gr\"ossen als ``Versor'' zu bezeichnen, w\"ahrend der Name ``Vektor'' im alten Sinne des Vierervektors für den Fall der Kovarianz verbleiben soll. Wir haben es eigentlich mit der Erweiterung des rein geometrisch aufgefassten Begriffes der ``Strecke'' zu tun. Auch der Versor kann durch eine Strecke abgebildet werden, welche jedoch bei Drehung des Achsenkreuzes seine Richtung im allgemeinen nicht beibeh\"alt, sondern auch eine bestimmte Drehung erf\"ahrt. Ausserdem sollen die Komponenten auch komplex sein dürfen.

   Als grundlegende Operation führen wir statt der skalaren und vektoriellen Multiplikation einzig allein die Quaternionenmultiplikation ein. Wir sahen, dass bei dieser Multiplikation, um einen Versor zu erhalten, der Raumteil des einen Vektors mit negativem Vorzeichen zu nehmen ist. Das soll als ``Konjugierte'' des Vektors (oder der Quaternion) benannt und mit oben Strich bezeichnet werden --- in Analogie zu den komplexen Zahlen. Also : 
\begin{equation}
\mathcal{F}=( \, X,Y,Z,T \, ) \qquad \qquad \CON{\mathcal{F}}=(-X,-Y,-Z,T \, )
\end{equation}
Endlich sollen die nach den einzelnen Achsen zeigenden Einheitsvektoren mit den Symbolen $1_x , 1_y , 1_z$ und $1_t$ bezeichnet werden, durch sie wird ein Vektor folgendermasser dargestellt:
\begin{equation}
\mathcal{F}= X \, 1_x + Y \, 1_y + Z \, 1_z + T \, 1_t
\end{equation}
Die Rechenregeln für die in die Zeitachse fallende Einheit sind mit jenen für die gew\"ohnliche Einheit geltenden identisch, so dass auch:
\begin{equation}
1_t = 1
\end{equation}
gesetzt werden kann.

   Im übrigen gelten für die Multiplikation bekanntlich das distributive wie auch das assoziative Gesetz, w\"ahrend die Regel der Kommutation in folgender Gleichung ihren Ausdruck findet:\footnote{Es sei an dieser Stelle die bemerkenswerte Tatsache erw\"ahnt, dass alle Regeln der Multiplikation auch für quadratische Matrizen, insbesondere orthogonale gelten, wobei als Conjugierte einer Matrix die durch Versetzung der Horizontalen zur Vertikalen entstehende Matrix zu verstehen ist.}
\begin{equation}
\mathcal{F} \, \mathcal{G} = \CON{\CON{\mathcal{G}} \, \CON{\mathcal{F}}}
\end{equation}

   Das Produkt $\mathcal{F} \CON{\mathcal{F}}$ --- ein reiner Zeitversor, welcher auch als blosse Zahl angesehen werden kann --- stellt das Quadrat von der L\"ange des Vektors vor. Daraus ist sofort auch die Division abzuleiten. Der Quotient zweier Vektoren:
\begin{equation}
\mathcal{X} = \frac{\mathcal{F}}{\mathcal{G}}
\end{equation}
soll durch die Gleichung:
\begin{equation}
\mathcal{X} \mathcal{G} = \mathcal{F}
\end{equation}
bestimmt sein. Dann ist:
\begin{equation}
\mathcal{X} \mathcal{G} \CON{\mathcal{G}} = \mathcal{F} \, \CON{\mathcal{G}}
\end{equation}
also:
\begin{equation}
\mathcal{X} = \frac{\mathcal{F} \, \CON{\mathcal{G}}}{\mathcal{G} \, \CON{\mathcal{G}}} 
\end{equation}
Damit ist die Division auf eine Multiplikation und eine rein skalare Division zurückgeführt.

\chapter{Charakterisierung der vierdimensionalen
                Drehung durch\\ zwei Quaternionen.}

Es besteht ein merkwürdiger Zusammenhang zwischen den Quaternionen und der allgemeinen orthogonalen Transformation im vierdimensionalen Raum, welche eine einfache und naturgem\"asse Bestimmung einer beliebigen Drehung des Achs\-en\-kreuzes erm\"oglicht. Zu dieser Bestimmung sind 6 unabh\"angige Gr\"ossen erforderlich, da ja die 16 Koeffizienten der Transformation-Matrix den 10 Orthogonalit\"atsbe\-ding\-ungen entsprechen müssen.

   Nehmen wir eine Quaternion von der L\"ange Eins:	
\begin{equation}
p=( \, p_1 , p_2 , p_3 , p_4 \, ) 
\end{equation}
und multiplizieren wir sie mit dem Vektor:
\begin{equation}
\mathcal{F}=( \, X , Y , Z , T \, ) 
\end{equation}
Die Komponenten des Produktes sind:
\begin{align}
\left.
\begin{array}{c}
X'   =   + p_4 X - p_3 Y + p_2 Z + p_1 T  \\
Y'   =   + p_3 X + p_4 Y - p_1 Z + p_2 T  \\
Z'   =   - p_2 X + p_1 Y + p_4 Z + p_3 T  \\
T'   =   - p_1 X - p_2 Y - p_3 Z + p_4 T 
\end{array}
\quad \right\}
\end{align}
Fassen wir das als eine Transformation des Vektors $\mathcal{F}$ in $\mathcal{F}'$ auf, so sehen wir, dass wir es mit einer Drehung zu tun haben, denn die Matrix der Transformation:
\begin{equation}
\mathcal{P}=
\begin{array}{|rrrr|}
p_4 & -p_3 & p_2 & p_1  \\
p_3 & p_4 & -p_1 & p_2  \\
-p_2 & p_1 & p_4 & p_3  \\
-p_1 & -p_2 & -p_3 & p_4 
\end{array}
\end{equation}
--- sie geh\"ort zu der sogenannten antisymmetrischen Matrizen --- ist orthogonal. Dasselbe ist der Fall, wenn die Quaternion den zweiten Faktor bildet. Dann ist die Matrix --- für welche der Buchstabe $\mathcal{Q}$ gebraucht werden soll:
\begin{equation}
\mathcal{Q}=
\begin{array}{|rrrr|}
q_4 & q_3 & -q_2 & q_1  \\
-q_3 & q_4 & q_1 & q_2  \\
q_2 & -q_1 & q_4 & q_3  \\
-q_1 & -q_2 & -q_3 & q_4 
\end{array}
\end{equation}
Diese beiden Typen der Transformation wollen wir der Kürze halber als $\mathcal{P}$-Transformation, bzw., $\mathcal{Q}$-Transformation, die beiden Matrizen als $\mathcal{P}$- und $\mathcal{Q}$-Matrizen bezeichnen. Die $\mathcal{P}$- und $\mathcal{Q}$-Transformationen bilden jede für sich Untergruppen in der allgemeinen Gruppe der orthogonalen Transformationen, das heisst, zwei nacheinander ausgeführte $\mathcal{P}$-Transformation führen wieder zu einer $\mathcal{P}$-Transformation und entsprechendes gilt für die $\mathcal{Q}$-Transformation. Das folgt aus dem assoziativen Gesetz der Multiplikation. Es sei n\"amlich:
\begin{equation}
\left.
\begin{array}{c}
\mathcal{F}=p_1 \, \mathcal{F}' \\
\mathcal{F}'=p_2 \, \mathcal{F}''
\end{array}
\quad \right\}
\end{equation}
dann ist:
\begin{equation}
\mathcal{F}=(p_1 p_2)\mathcal{F}''
\end{equation}
Andererseits sei:
\begin{equation}
\left.
\begin{array}{c}
\mathcal{F} = \mathcal{F}' q_1 \\
\mathcal{F}' = \mathcal{F}'' q_2
\end{array}
\quad \right\}
\end{equation}
dann ist:
\begin{equation}
\mathcal{F}=\mathcal{F}''(q_2 q_1)
\end{equation}
Zu jeder $\mathcal{P}$- oder $\mathcal{Q}$-Matrix geh\"ort eine Quaternion. Schreiben wir dieselbe als Index, so gelten also nach den eben hingeschriebenen Gleichungen für die Produkte zweier $\mathcal{P}$- bzw., $\mathcal{Q}$-Matrizen die Regeln:
\begin{equation}
\left.
\begin{array}{c}
\mathcal{P}_{p_1} \, \mathcal{P}_{p_2} =\mathcal{P}_{p_1 \, p_2} \\
\mathcal{Q}_{q_1} \, \mathcal{Q}_{q_2} =\mathcal{Q}_{q_2 \, q_1}
\end{array}
\quad \right\}
\end{equation}

   Führen wir nun nach einer $\mathcal{P}$-Transformation eine $\mathcal{Q}$-Transformation aus, so erhalten wir wieder eine orthogonale Transformation, und zwar --- wie sich zeigen l\"asst --- die allgemeine. Die Zusammensetzung der $\mathcal{P}$- und $\mathcal{Q}$-Gruppen ergibt also die Gruppe der allgemeinen orthogonalen Transformation. Da die Richtung der beiden Quaternionen $p$ und $q$ beliebig gew\"ahlt werden kann, w\"ahrend ihre L\"ange die Einheit sein muss, so verfügen wir tats\"achlich über 6 Glieder. Die Reihenfolge der beiden Transformationen ist übrigens gleichgültig, es gilt hier das kommutative Gesetz:
\begin{equation}
\mathcal{P} \, \mathcal{Q} = \mathcal{Q} \, \mathcal{P}
\end{equation}
Die allgemeine orthogonale Transformation des Vektors $\mathcal{F}$ in $\mathcal{F}'$ kann somit durch die einfache Gleichung:
\begin{equation}
\mathcal{F}' = p \, \mathcal{F} \,q
\end{equation}
wiedergegeben werden. Jede orthogonale Matrix ist als Produkt einer $\mathcal{P}$- und $\mathcal{Q}$-Matrix darstellbar, und zwar sind die zugeh\"origen Quaternionen $p$ und $q$ (abgesehen von einer Multiplikation mit $-1$) eindeutig bestimmt. Man kann sie als die Charakteristiken der Transformation bezeichnen.

   Die resultierende Matrix wird durch die Charakteristiken sehr einfach darge\-stel\-lt. Wir schreiben die Komponenten der Produkte:
\begin{equation}
p \, 1_x \, q \quad , \quad p \, 1_y \, q \quad , \quad p \, 1_z \, q \quad , \quad p \, q 
\end{equation}
in je eine Vertikale unter einander, in der gewohnten Reihenfolge, diese 16 Komponenten ergeben die 16 Koeffizienten der orthogonalen Matrix. Es sei, z.B., die allgemeine orthogonale Matrix folgendemassen bezeichnet:
\begin{equation}
\begin{array}{c|cccc|}
~  & x & y & z & t  \\
\hline
x' & \alpha_{11} & \alpha_{12} & \alpha_{13} & \alpha_{14}  \\
y' & \alpha_{21} & \alpha_{22} & \alpha_{23} & \alpha_{24}  \\
z' & \alpha_{31} & \alpha_{32} & \alpha_{33} & \alpha_{34}  \\
t' & \alpha_{41} & \alpha_{42} & \alpha_{43} & \alpha_{44}   
\end{array}
\end{equation}
Dann ist:
\begin{equation}
\alpha_{11} \, 1_x + \alpha_{21} \, 1_y +  \alpha_{31} \, 1_x +  \alpha_{41} = p \, 1_x \, q
\end{equation}
und so weiter.  Mann kann auch so verfahren, dass man die entsprechenden Komponenten der Produkte:
\begin{equation}
\CON{p} \, 1_x \, \CON{q} \quad , \quad \CON{p} \, 1_y \, \CON{q} \quad , \quad \CON{p} \, 1_z \, \CON{q} \quad , \quad \CON{p} \,   \, \CON{q} 
\end{equation}
in je eine $\emph{Horizontale}$ schreibt.

   Ist es umgekehrt die Aufgabe, zu einer gegebenen Matrix die Charakteristiken zu suchen, so gehen wir in symmetrischer Weise folgendermassen vor. Die Vertikalen seien als Quaternionen betrachtet der Reihe nach: $\mathcal{U}_1,\mathcal{U}_2, \mathcal{U}_3, \mathcal{U}_4$.  Dann ist:
\begin{gather}
\left.
\begin{array}{c}
p = \frac{-1}{4\lambda} (\, \mathcal{U}_1 \, 1_x + \mathcal{U}_2 \, 1_j + \mathcal{U}_3 \, 1_z - \mathcal{U}_4 \,) \\
~\\
q = \frac{-1}{4\mu} (\, 1_x \, \mathcal{U}_1 + 1_j \, \mathcal{U}_2 + 1_z  \, \mathcal{U}_3 - \mathcal{U}_4 \,)
\end{array}
\quad \right\}
\end{gather}
$\lambda$ und $\mu$ bedeuten blosse Zahlen. Sie werden (bis auf der Faktor $\pm 1$) durch die Forderung bestimmt, dass die L\"ange von $p$ und $q$ gleich Eins, ausserdem:
\begin{equation}
\lambda \, \mu = \frac{1}{4} (\alpha_{11} + \alpha_{22} +\alpha_{33} +\alpha_{44})
\end{equation}
sein muss.

   Bei einer rein r\"aumlichen Transformation ist die vierte Vertikale als Quaternion betrachtet:
\begin{equation}
\mathcal{U}_4 = 1 
\end{equation}
Es ist also dann : $p \, q=1$, oder:
\begin{equation}
q = \CON{p} 
\end{equation}
Die Untergruppe der rein r\"aumlichen Transformationen ist somit dadurch definierbar, dass alsdann die beiden Charakteristiken zu einander konjugiert sind.

   Wir haben im vorigen Kapitel die Versoren eingeführt und sahen, dass das Produkt $\mathcal{F}\,\CON{\mathcal{G}}$ aus den Vektoren $\mathcal{F}$ und $\mathcal{G}$ einen solchen Versor darstellt. Jetzt wollen wir noch nachtr\"aglich die Matrix seiner Transformation bestimmen. Es sei die orthogonale Transformation durch:
\begin{gather}
\left.
\begin{array}{c}
\mathcal{F}' = p \, \mathcal{F} \, q \\
\mathcal{G}' = p \, \mathcal{G} \, q
\end{array}
\quad \right\}
\end{gather}
gegeben. Nach den Regeln der Multiplikation ist:
\begin{equation}
\CON{\mathcal{G}'} = \CON{q} \, \CON{\mathcal{G}} \, \CON{p}
\end{equation}
und somit ist:
\begin{equation}
\mathcal{F}' \, \CON{\mathcal{G}'} = p \, \mathcal{F} \, \CON{\mathcal{G}} \, \CON{p} 
\end{equation}
Die Matrix der Transformation wird also durch das Produkt:
\begin{equation}
\mathcal{P}_p \, \mathcal{Q}_{\CON{p} }
\end{equation}
dargestellt, aber auch durch:
\begin{equation}
\mathcal{P}_p \, \mathcal{Q}_q \, \mathcal{Q}_{\CON{q} }  \,  \mathcal{Q}_{\CON{p} } = (\mathcal{P}_p \, \mathcal{Q}_q) (\mathcal{Q}_{\CON{p} \, \CON{q} })
\end{equation}
Ausführlicher hingeschrieben --- in Betracht genommen, dass $ \CON{p} \, \CON{q} $ die vierte Horizontale der Vektormatrix darstellt:
\begin{gather}
\begin{array}{|cccc|}
\alpha_{11} & . & . & \alpha_{11}  \\
. & . & . & .  \\
. & . & . & .  \\
\alpha_{41} & . & . & \alpha_{44} 
\end{array}
~~
\begin{array}{|rrrr|}
~\alpha_{44} & ~\alpha_{43} & -\alpha_{42} & \alpha_{41}  \\
-\alpha_{43} & ~\alpha_{44} & ~\alpha_{41} & \alpha_{42}  \\
~\alpha_{42}& -\alpha_{41} & ~\alpha_{44} & \alpha_{43} \\
-\alpha_{41} & -\alpha_{42} & -\alpha_{43} & \alpha_{44} 
\end{array}
\end{gather}
Die erste Matrix ist die Vektormatrix selber. Die Transformation des Produktes $\mathcal{F} \, \CON{\mathcal{G}}$ ergibt sich somit als Resultante zweier orthogonalen Transformationen. Die erste ist die Transformation der Faktoren $\mathcal{F}$ und $\mathcal{G}$, die zweite eine bestimmte $\mathcal{Q}$-Transformation.  Würd also der Versor $\mathcal{F} \, \CON {\mathcal{G}}$ durch eine Strecke abgebildet, so erf\"ahrt diese Strecke bei Drehung des Koordinatensystems eine durch die zweite Matrix gegebene Drehung. Diese unterbleibt nur im Falle einer rein r\"aumlichen Transformation (dann ist die zweite Matrix gleich Eins), alsdann geht der Versor in einen gew\"ohnlichen Vektor über. \"{A}hnlich liegen die Verh\"altnisse auch beim Produkt: $\CON{\mathcal{F}} \, \mathcal{G}$\,.

   In der physikalischen Anwendung kommt die vierdimensionale orthogonale Transformation in Form der L{\scriptsize ORENTZ}-Transformation in Betracht, wo die r\"aumli\-chen Koordinaten reelle Gr\"ossen sind, die Zeitkoordinate hingegen imagin\"ar. Die Koeffizienten der Transformation sind demgem\"ass zum Teil reell, zum Teil rein imagin\"ar. Die beiden Charakteristiken $p$ und $q$ stellen aber dann komplexe Gr\"ossen vor. Wir ben\"otigen hier noch eine Bezeichnung, n\"amlich um den konjugiert komplexen Wert der komplexen Quaternion:
\begin{equation}
p = p' + i p'' 
\end{equation}
ausdrücken zu k\"onnen. Da das gewohnte Zeichen $\CON{(~)}$ schon in anderer Bedeutung besetzt ist, soll hier für das \"ahnliche Symbol $(~)^*$:
\begin{equation}
p^* = p' -i p''
\end{equation}
dienen. Es sei nun die L{\scriptsize ORENTZ}-Transformation durch die Gleichung:
\begin{equation}
\mathcal{F}'= p \, \mathcal{F} \, q
\end{equation}
ausgedrückt. Setzen wir statt $i$ überall $-i$, so wird:
\begin{equation}
\mathcal{F}'{}^*= p^* \, \mathcal{F}^* \, q^*
\end{equation}
Hat der Vektor die Eigenschaft, dass sein Raumteil rein reell, sein Zeitteil rein imagin\"ar ist, so beh\"alt er (im Falle der L{\scriptsize ORENTZ}-Transformation) diese Eigenschaft auch nach der Transformation bei. Daraus folgt aber, dass zwischen $\mathcal{F}'{}^*$ und $\mathcal{F}^*$ dieselbe Beziehung bestehen muss, wie zwischen $\CON{\mathcal{F}'}$ und $\CON{\mathcal{F}}$. Es ist aber:
\begin{equation}
\CON{\mathcal{F}'}= \CON{q} \, \CON{\mathcal{F}} \, \CON{p}
\end{equation}
und durch Vergleich mit der früheren Gleichung folgt hieraus:
\begin{equation}
                       \CON{q} = p^*
\end{equation}
und
\begin{equation}
                       \CON{p} = q^*
\end{equation}
Diese zwei Formeln sind miteinander identisch.

   Wir sehen also im Falle der L{\scriptsize ORENTZ}-Transformation ist die Konjugierte der einen Charakteristik gleich dem konjugiert komplexen Werte der anderen Reell sind die Charakteristiken nur bei rein r\"aumlicher Drehung, denn alsdann ist: $\CON{q}=p$, also auch:
\begin{equation}
p^* = p
\end{equation}

\noindent{---------------------}

(\emph{Anmerkung}) Am Schlusse dieses Kapitels, welches mit dem vorigen beisammen die formale Grundlage der folgenden Entwicklungen enth\"alt, m\"ochte ich noch kurz erw\"ahnen, dass die Quaternionenmultiplikation auch hinsichtlich der Tensoren --- als ein quadratisches Schema, welches die Transformation eines Vektors in wiederum einen Vektor bewirkt --- eine genügende Basis zu bilden scheint. Man gelangt n\"amlich aus den beiden Vektoren: $\mathcal{F}$ und $\mathcal{G}$ zu einem Tensorschema, wenn man die Komponenten von den vier Produkten:
\begin{equation}
-\mathcal{F} \, 1_x \, \mathcal{G} \quad , \quad -\mathcal{F} \, 1_y \, \mathcal{G} \quad , \quad -\mathcal{F} \, 1_z \, \mathcal{G} \quad , \quad \mathcal{F} \, \mathcal{G}
\end{equation}
in je eine Vertikale untereinander schreibt in \"ahnlicher Weise, wie wir es bei der Bildung der orthogonalen Matrix aus den Charakteristiken getan haben.

\chapter{Die Quaternionfunktionen.}

Hinsichtlich der Transformation ist der sogenannte H{\scriptsize AMILTON}sche Operator:
\begin{equation}
\nabla = \frac{\partial ~}{\partial x} 1_x + \frac{\partial ~}{\partial y} 1_y + \frac{\partial ~}{\partial z} 1_z + \frac{\partial ~}{\partial t} 
\end{equation}
einem Vektor aequivalent. Führen wir also die Multiplikation:
\begin{equation}
\CON{\mathcal{F}} \, \nabla = \frac{\partial  \CON{\mathcal{F}}}{\partial x} 1_x + \frac{\partial  \CON{\mathcal{F}}}{\partial y} 1_y + \frac{\partial  \CON{\mathcal{F}}}{\partial z} 1_z + \frac{\partial  \CON{\mathcal{F}}}{\partial t}
\end{equation}
aus, wobei $\mathcal{F}$ ein Vektor sein soll, so erhalten wir einen Versor. Seine Komponenten sind --- wenn die Komponenten von $\mathcal{F}$ mit $X,Y,Z,T$ bezeichnet werden --- :
\begin{gather}
\left.
\begin{array}{c}
\dfrac{\partial \, T}{\partial x} + \dfrac{\partial \, Z}{\partial y} - \dfrac{\partial \, Y}{\partial z} - \dfrac{\partial \, X}{\partial t}    \\
~\\
\dfrac{\partial \, T}{\partial y} + \dfrac{\partial \, X}{\partial z} - \dfrac{\partial \, Z}{\partial x} - \dfrac{\partial \, Y}{\partial t}    \\
~\\
\dfrac{\partial \, T}{\partial z} + \dfrac{\partial \, Y}{\partial x} - \dfrac{\partial \, X}{\partial y} - \dfrac{\partial \, Z}{\partial t}    \\
~\\
\dfrac{\partial \, X}{\partial x} + \dfrac{\partial \, Y}{\partial y} + \dfrac{\partial \, Z}{\partial z} + \dfrac{\partial \, T}{\partial t}   
\end{array}
\quad \right\} 
\end{gather}
Setzen wir diese Ausdrücke gleich Null, so bleibt das so erhaltene Gleichungssystem den Eigenschaften der Versoren zufolge auch bei jeder beliebigen Drehung des Koordinatensystems bestehen. Diese partiellen Differentialgleichungen definieren gewisse Funktionen der vier Variabelen $x,y,z,t$, --- die wir hier s\"amtlich als reell voraussetzen --- wir wollen ihnen den Namen ``Quaternionfunktionen'' geben, und zwar den Inbegriff der vier Werte $X, Y, Z, T,$ als $\emph{eine}$ Funktion betrachtet. Es besteht n\"amlich symbolisch zwischen ihnen und den Quaternionen derselbe Zusammenhang, wie zwischen den komplexen Funktionen und komplexen Zahlen. Sind doch symbolisch die C{\scriptsize AUCHY}-R{\scriptsize IEMANN}schen Differentialgleichungen in der Gleichung:
\begin{equation}
(u+iv) \bigl( \frac{\partial  ~}{\partial x}
          + i \frac{\partial  ~}{\partial y} \bigr) = 0
\end{equation}
enthalten. Die Analogie geht aber weit über das formale hinaus. Die Gleichungen:
\begin{equation}
\CON{\mathcal{F}} \, \nabla = 0 
\end{equation}
oder auch:
\begin{equation}
\CON{\nabla } \mathcal{F} =0 
\end{equation}
k\"onnen in mancher Hinsicht als Repr\"asentanten der funktionentheoretischen Gr\-undgleichungen im vierdimensionalen Raume angesehen werden. Allerdings gehen manche klassische Eigenschaften der komplexen Funktionen verloren, viele der grundlegenden bleiben aber erhalten, oder sind in entsprechender Weise übertrag\-bar. So vor allen Dingen der Zusammenhang mit dem --- hier vierdimensionalen --- Potential. Es ist n\"amlich:
\begin{equation}
\nabla (\CON{\nabla } \mathcal{F}) = (\nabla \CON{\nabla }) \mathcal{F} = \frac{\partial ^2 \mathcal{F}}{\partial x^2 } + \frac{\partial ^2 \mathcal{F}}{\partial y^2 } + \frac{\partial ^2 \mathcal{F}}{\partial z^2 } + \frac{\partial ^2 \mathcal{F}}{\partial t^2 } = 0
\end{equation}
das heisst, alle vier Komponenten einer Quaternionfunktion sind Potentiale (gen\"u\-gen der L{\scriptsize APLACE}schen Gleichung). Auch l\"asst sich jede Quaternionfunktion umgekehrt auf vier Potentialfunktionen zurückführen. Es sei n\"amlich $\mathcal{F}$ in folgender Weise ausgedrückt:
\begin{equation}
\mathcal{F} =\nabla \CON{\Phi}
\end{equation}
dann ist $\mathcal{F}$ eine Quaternionfunktion, wenn:
\begin{equation}
\CON{\nabla } \nabla \, \CON{\Phi} =0 
\end{equation}
also die Komponenten von $\Phi$ Potentiale sind damit ist die L\"osung der Grundgleichungen auf die L\"osung der L{\scriptsize APLACE}schen Gleichung zurückgeführt (``Vektorpotential''). Im folgenden soll der Ausdruck: ``das Potential'' die Viererfunktion $\Phi$ bedeuten, aus welcher sich die Quaternionfunktion in der eben erw\"ahnten Weise ableitet.

   Auch der klassische fundamentale C{\scriptsize AUCHY}sche Integralsatz, welcher die Bestimmung der komplexen Funktion aus den Randwerten erlaubt, findet sein vollkommenes Analogon. Sein Beweis geschieht ganz \"ahnlich wie in der Funktionentheorie. Wir wenden den G{\scriptsize AUSS}schen Integralsatz auf die Gleichung:
\begin{equation}
\CON{\mathcal{F}} \, \nabla = \frac{\partial  \CON{\mathcal{F}}}{\partial x} \, 1_x + \frac{\partial  \CON{\mathcal{F}}}{\partial y} \, 1_y + \frac{\partial  \CON{\mathcal{F}}}{\partial z} \, 1_z + \frac{\partial  \CON{\mathcal{F}}}{\partial t}=0
\end{equation}
an und schreiben diese in der Form:
\begin{equation}
\int \CON{\mathcal{F}} \, n \, df=0
\end{equation}
dabei bedeute $\emph{n}$ die nach aussen weisende Fl\"achennormale, als Vektor genommen mit der L\"ange Eins, $df$ ist das Fl\"achenelement und die Integration soll über eine beliebige, jedoch lauter regul\"are Punkte umschliessende geschlossene Fl\"ache (im vierdimensionalen Raum) ausgestreckt werden. Weiterhin sei für eine andere Funktion $\mathcal{G}$:
\begin{equation}
                 \nabla \, \CON{\mathcal{G}} = 0
\end{equation}
also:      
\begin{equation}
1_x \, \frac{\partial \, \CON{\mathcal{G}}}{\partial x } + 1_y  \, \frac{\partial \, \CON{\mathcal{G}}}{\partial y } + 1_z  \, \frac{\partial \, \CON{\mathcal{G}}}{\partial z } +  \, \frac{\partial   \, \CON{\mathcal{G}}}{\partial t } = 0 
\end{equation}
Aus diesen zwei Gleichungen für $\mathcal{F}$ und $\mathcal{G}$ resultiert:
\begin{equation}
\frac{\partial  ~}{\partial x } (\CON{\mathcal{F}} \, 1_x \, \CON{\mathcal{G}}) + \frac{\partial  ~}{\partial y } (\CON{\mathcal{F}} \, 1_y \, \CON{\mathcal{G}}) + \frac{\partial  ~}{\partial z } (\CON{\mathcal{F}} \, 1_z \, \CON{\mathcal{G}}) + \frac{\partial  ~}{\partial t } (\CON{\mathcal{F}} \, \CON{\mathcal{G}}) = 0 
\end{equation}
oder mit Anwendung des G{\scriptsize AUSS}schen Satzes:
\begin{equation}
\int \CON{\mathcal{F}} \, n \, \CON{\mathcal{G}} \, df=0
\end{equation}
Das Innere der Fl\"ache muss hierbei sowohl in Bezug auf $\mathcal{F}$, wie in Bezug auf $\mathcal{G}$ regul\"ar sein. Nun w\"ahlen wir für $\mathcal{G}$ die Funktion:
\begin{equation}
\mathcal{G} = \nabla \frac{1}{R^2}
\end{equation}
\begin{equation}
R^2 = ( x - \xi )^2 + ( y - \eta )^2 + ( z - \zeta )^2 +( t - \vartheta)^2
\end{equation}
Sie hat einen einzigen singul\"aren Punkt, n\"amlich den Punkt:
\begin{equation}
x = \xi \quad , \quad y = \eta \quad , \quad z = \zeta \quad , \quad t = \vartheta
\end{equation}
Diesen schalten wir durch eine herumgeschlagene Kugel aus dem Integrationsgebiet aus und gelangen schliesslich durch \"ahnliche Folgerungen wie bei den komplexen Funktionen zur Gleichung:
\begin{equation}
\CON{\mathcal{F}}(\xi , \eta , \zeta , \vartheta) = \frac{-1~}{4 \pi^2} \int \CON{\mathcal{F}}(x,y,z,t) \, n \: \CON{\nabla} \frac{1}{R^2} \, df
\end{equation}
wo die Integration rechterhand auf eine, den Punkt $( \, \xi , \eta , \zeta , \vartheta \, )$ umschliessende, im Inneren überall regul\"are Fl\"ache zu erstrecken ist. Diese Gleichung bestimmt die Werte der Funktion im Inneren eines regul\"aren Raumgebietes durch die Randwerte.

   Die Analogie zum C{\scriptsize AUCHY}schen Satze ist auffallend, besonders, wenn letzterer in entsprechender Form geschrieben wird. Denn man kann statt der gewohnten Form:
\begin{equation}
f(\xi + i\eta) = \frac{1}{2 \pi i } \int \frac{f(x+iy)}{(x-\xi)+i(y-\eta)} d(x+iy)
\end{equation}
auch setzen:
\begin{equation}
f(\xi + i\eta) = \frac{1}{4 \pi} \int f(x+iy) \, n \, \CON{\nabla} \, ( \, \log R^2 \, ) \, ds
\end{equation}
$ds$ bedeutet das Linienelement, $n$ und $R$ das entsprechende, wie in der Formel für $\mathcal{F}$,
\begin{equation}
\CON{\nabla} = \frac{\partial  ~}{\partial x } -i\frac{\partial  ~}{\partial y }
\end{equation}
Dem logarithmischen Potential: $\log \, R^2$ entspricht in vier Dimensionen als Potential: $1/R^2$.

   Aus dem abgeleiteten Integralsatz lassen sich hinsichtlich der Reihenentwicklung nach auf- und absteigenden Potenzen der Entfernung (``L{\scriptsize AURENT}sche Reihe'') für die Quaternionfunktionen \"ahnliche Resultate ableiten, welche für die komplexen Funktionen gültig sind, nur dass hier für die Multiplikation das kommutative Gesetz nicht erfüllt ist. Diese Er\"orterungen, da sie in erster Linie von rein mathematischem Interesse sind, sollen hier übergangen werden.

\chapter{Die Maxwellschen Gleichungen.}

Wenn die Komponenten der Funktion $\mathcal{F}$ komplexe Gr\"ossen sind und die vierte Variabele imagin\"ar:
\begin{equation}
                          t = i \tau
\end{equation}
wo $\tau$ die Zeit bedeutet (in einem Massystem mit der Lichtgeschwindigkeit $c=1$) so gehen die Definitionsgleichungen der Quaternionfunktionen in die M{\scriptsize AXWELL}\-schen Gleichungen über und zwar in dem speziellen Fall, wenn wir $T=0$ setzen. Es besteht also ein \"ausserst inniger Zusammenhang zwischen den Grundgleichungen des Elektromagnetismus und den im vorigen Kapitel eingeführten Quaternionfunktionen. Die elektrische und magnetische Feldst\"arke bildet dabei in dieser Auffassung keinen Sechservektor, sondern einen komplexen Versor, in der Form:
\begin{equation}
                    \mathcal{F} = \mathcal{H} + i\mathcal{E}
\end{equation}
wenn $\mathcal{E}$ die elektrische, $\mathcal{H}$ die magnetische Feldst\"arke bedeutet. In der gewohnten dreidimensionalen Symbolik geschrieben bestehen n\"amlich zwischen den beiden Feldst\"arken folgende Beziehungen:
\begin{gather}
\left.
\begin{array}{c}
\dfrac{\partial \mathcal{H}}{\partial \tau }
     + \operatorname{curl} \,\mathcal{E} = 0  \\ ~\\
\dfrac{\partial \mathcal{E}}{\partial \tau }
     - \operatorname{curl} \,\mathcal{H} = 0
\end{array}
\quad \right\}
\end{gather}
\begin{gather}
\left.
\begin{array}{c}
\operatorname{div} \, \mathcal{E} = 0 \\
~\\
\operatorname{div} \, \mathcal{H} = 0
\end{array}
\quad \right\}
\end{gather}
also:
\begin{gather}
\left.
\begin{array}{c}
           \dfrac{\partial  (\mathcal{H}+i\mathcal{E})}{i \partial \tau }
     \operatorname{curl} \, (\mathcal{H}+i\mathcal{E})  = 0 \\
~\\
     \operatorname{div}  \, (\mathcal{H}+i\mathcal{E})  = 0
\end{array}
\quad \right\}
\end{gather}
oder:
\begin{gather}
\left.
\begin{array}{c}
            \dfrac{\partial \mathcal{F}}{\partial t}
    - \operatorname{curl} \, \mathcal{F} = 0 \\
~\\
      \operatorname{div}  \, \mathcal{F} = 0
\end{array}
\quad \right\}
\end{gather}
Die Gleichung:
\begin{equation}
\CON{\nabla} \, \mathcal{F} = 0
\end{equation}
vereinigt beide Gleichungsysteme für $T=0$. Von dieser Beschr\"ankung wollen wir uns jedoch prinzipiell freimachen, da sie, von dem hier eingenommenen Standpunkte aus betrachtet, als unbegründete Voraussetzung der ganzen Natur der Untersuchung widersprechen würde.

   An dieser Stelle m\"ochten wir noch die Frage der Transformation für die Funktion $\mathcal{F}$ berühren, falls wir aus einem gegebenen Koordinatensystem zu einem \"aquivalenten übergehen. Wir sahen, dass die Grundgleichungen invariant bleiben, wenn $\mathcal{F}$ als Vektor betrachtet wird. Nun bildet aber in der M{\scriptsize INKOWSKI}schen Auffassung die elektrische und magnetische Feldst\"arke zusammen einen Sechservektor, befolgt also ganz andere Transformationsformeln. Der Sachverhalt ist so, dass die Transformation des Versors $\mathcal{F}$ überhaupt nicht eindeutig festgelegt ist. Denn es ist leicht einzusehen, dass mit $\mathcal{F}$ zugleich auch:
\begin{equation}
\mathcal{F} ' = \mathcal{F} \, q_0
\end{equation}
eine Quaternionfunktion ist, falls $q_0$ eine konstante Quaternionen bedeutet, deren L\"ange wir $|q_0| = 1$ w\"ahlen, um eine blosse \"{A}hnlichkeitstransformation im vorhin\-ein auszuschalten. Diese Gleichung bedeutet aber eine orthogonale Transformation für die Funktion $\mathcal{F}$ und zwar eine $\mathcal{Q}$-Transformation. Wir sehen also, dass schon in einem und demselbe Koordinatensystem eine dreidimensionale Mannigfaltigkeit von Transformationen für die Komponenten einer Quaternionfunktion besteht, welche die bestimmenden Gleichungen unberührt lassen. Führen wir nun eine Drehung des Achsenkreuzes mit den Charakteristiken $p$ und $q$ aus, so wird die Transformation des Versors $\mathcal{F}$ im allgemeinen durch die Formel:
\begin{equation}
\mathcal{F} ' = p \, \mathcal{F} \, q \, q_0
\end{equation}
wiedergegeben, d.h., nach der Vektortransformation kann eine beliebige $\mathcal{Q}$-Trans\-formation mit der Charakteristik $q_0$ noch ausgeführt werden. W\"ahlen wir insbesondere:
\begin{equation}
q_0 = \CON{q} \, \CON{p}
\end{equation}
ist also die Formel der Transformation:
\begin{equation}
\mathcal{F} ' = p \, \mathcal{F} \, \CON{p}
\end{equation}
so haben wir die im elektromagnetischen Felde übliche Transformation vor uns. Der zuerst erw\"ahnte Fall, dass $\mathcal{F}$ unmittelbar ein Vektor ist ($q_0=1$), liegt im Gravitationsfelde vor.  Hier leitet sich die Feldst\"arke aus einem skalaren Potential (Zeitkomponente des allgemeinen Potentials $\Phi$) ab, als dessen Gradient. Das Potential bleibt bei einer Drehung invariant, die Feldst\"arke transformiert sich also als Vektor.

   Die Transformation des Potentials $\Phi$ wird im allgemeinen durch die Formel angegeben:
\begin{equation}
\Phi ' = \CON{q_0} \, \CON{q} \, \Phi \, q 
\end{equation}
Führen wir für $q_0$ seinen Wert im elektromagnetischen Felde ein, so resultiert für $\Phi $ die Vektortransformation. W\"ahrend also im Gravitationsfelde die Feldst\"arke ein Vektor ist, ist es im elektromagnetischen das Potential (``Vektorpotential''). Auf alle F\"alle representieren sowohl Feldst\"arke, wie Potential, einen Versor.

\chapter{Das Elektron als funktionentheoretische Singularit\"at.}

Die Grundgleichungen der M{\scriptsize AXWELL}schen Theorie sollen in ihrer ursprünglichen erhabenen Einfachheit und Unwillkürlichkeit nur für den reinen Aether Gültigkeit besitzen, für die Materie werden sie mit neuen fremden Gliedern erg\"anzt. Dieses Verfahren steht in einem merkwürdigen Kontrast zu dem wirklichen mathematischen Geiste dieser Gleichungen. Wir haben es mit partiellen Differentialgleichungen zu tun, und was das bedeutet, was für eine Fülle der M\"oglichkeiten schon hinter den einfachsten Typen verborgen liegt, dafür gibt es ein klassiches Beispiel in der Funktionentheorie, welche ihr grossartiges Geb\"aude allein auf Grund der C{\scriptsize AUCHY}-R{\scriptsize IEMANN}schen Differentialgleichungen aufbaut. Ob auch auf die Quaternionfunktionen eine entsprechend weitschichtige Disziplin zu gründen w\"are, kann fraglich sein, sicher aber ist, dass die Funktionentheorie den Weg, die Methode, die allgemeinen Gesichtspunkte anweist, wie partielle Differentialgleichungen überhaupt anzufassen sind. Von einer Erg\"anzung mit fremden, nicht zur Funktion geh\"orenden Gliedern, kann dabei nicht die Rede sein. Es gibt wohl Stellen, wo die Gleichungen ihre Gültigkeit verlieren, diese sind jedoch nicht willkürlich, sondern durch die Natur der Funktion selbst bestimmt. An solchen Stellen kann überhaupt von Gleichungen nicht mehr gesprochen werden, weil die Differentialquotienten ihre Bedeutung verlieren. Man pflegt sie im Gegensatz zu den regul\"aren als singul\"are Stellen der Funktion zu bezeichnen. Sie sind derart charakteristisch, dass sie durch ihre Lage und das Verhalten der Funktion in ihrer unmittelbaren Umgebung eine naturgem\"asse Bestimmung der betreffenden Funktion erm\"oglichen. Durch eine solche funktionentheoretische Deutung --- die übrigens schon durch die im vorigen Kapitel berührte merkwürdige Verwandtschaft der hier in Rede stehenden Funktionen mit den komplexen nahegelegt wird --- erh\"alt das Problem der Materie, insbesondere ihre atomistische Struktur, eine ausserordentlich harmonische L\"osung. \emph{Die Materie repr\"asentiert die singul\"aren Stellen derjenigen Funktionen, welche durch die im Aether gültigen Differentialgleichungen bestimmt werden.}

   Die vorhin erw\"ahnte charakteristische Rolle der Singularit\"aten findet ihren physikalischen Sinn in der Tatsache, dass alle Wirkungen ihren Ausgangspunkt in der Materie haben. Die Grundparadoxie der Elektronentheorie: wie ein Gebilde bei lauter abstossenden Kr\"aften zusammengehalten werden kann, wird hier gegenstandslos, w\"ahrend die Diskontinuit\"at der Materie als selbstverst\"andliche Konsequenz erscheint, da ja die singul\"aren Punkte naturgem\"ass eine diskrete Menge im regul\"aren Raume bilden. Eine konsequent durchgeführte Feldtheorie, welche im Geiste der partiellen Differentialgleichungen aufgebaut werden will, steht also nicht nur in keinem Widerspruch zum Atomismus, sodern führt sogar direkt zu ihn hin.

   Nun bilden aber die Grundgleichungen ein lineares und homogenes System, demzufolge beliebige partikul\"are L\"osungen ganz unabh\"angig übereinander gelagert werden k\"onnen. Die Lage der einzelnen Singularit\"aten, ebenso wie die ihnen angeh\"orenden Funktionen als partikul\"are L\"osungen, sind von einander vollkommen unabh\"angig. In Wirklichkeit finden wir nach beiden Punkten hin, einerseits fortw\"ahrende Wechselwirkung, andererseits strenge Determiniertheit. Die Grundgleichungen k\"onnen also zur Erkl\"arung der Natur nicht ausreichen, sie müssen noch durch ein neues Prinzip erg\"anzt werden. Zur L\"osung dieser Frage m\"ochte das n\"achste Kapitel einen Beitrag liefern.

\chapter{Das Hamiltonsche Prinzip.}

Von nun an wollen wir die vierte Variabele als rein imagin\"ar voraussetzen, welche mit der Zeit $\tau$ durch die Gleichung:
\begin{equation}
                    t = i \tau
\end{equation}
zusammenh\"angen soll. Dann ist die grundlegende, partikul\"are L\"osung der Potentialgleichung die folgende Funktion:
\begin{equation}
                   \Phi = \frac{ f(\tau') \sqrt{1-v^2} }{ r ( 1 + \dot{r} )}
                   \qquad ,\qquad
                   v^2 = \dot{\xi}^2 + \dot{\eta}^2 + \dot{\zeta}^2 ~~.
\end{equation}
Hier haben wir es mit einer Singularit\"at zu tun, welche im $( \, x, y, z, \tau  \, )$ Raume abgebildet eine Linie darstellt (``Weltlinie''). Die r\"aumlichen Koordinaten seiner Punkte: $\xi, \eta, \zeta$, k\"onnen beliebige Funktionen der Zeit sein, (nur darf keine \"{U}berlichtgeschwindigkeit eintreten,) $v$ ist die Geschwindigkeit, $\tau'$ die retardierte Zeit:
\begin{equation}
                         \tau' = \tau - r
\end{equation}
wobei $r$: die Entfernung des Aufpunktes $(\, x, y, z \,)$ von $(\, \xi, \eta, \zeta \,)$, jedoch im Zeitpunkte $\tau'$ genommen bedeutet. Die Punkte bedeuten Differentiationen nach ebenfalls $\tau'$.

   Aus dieser L\"osung lassen sich durch Differentiation nach den Koordinaten des Aufpunktes: $( \, x, y, z, \tau \, )$ neue partikul\"are L\"osungen bilden, dann nach demselben Verfahren wiederum neue usw., und durch Zusammensetzung all dieser partikul\"aren L\"osungen erhalten wir schliesslich eine unendliche Reihe nach absteigenden Potenzen der Entfernung, welche die allgemeine L\"osung der Potenzialgleichung für die betreffende Singularit\"at bildet. Diese Reihe (welche hier der L{\scriptsize AURENT}schen Entwicklung in der Funktionentheorie entspricht) enth\"alt als Koeffizienten eine unendliche Folge von willkürlichen Funktionen der Zeit.

   Ein Elektron kann somit als ein Gebilde mit unendlich vielen Freiheitsgraden angesehen werden.  Je genauer der Wert der Funktion bekannt sein muss, desto mehr Glieder der Reihe, also auch desto mehr Freiheitsgrade müssen berücksichtigt werden. In geh\"origer Entfernung jedoch wird der Einfluss der h\"oheren Glieder gegenüber den ersten immer geringer, und schliesslich kommen wir in eine Entfernung, wo wir die Reihe schon beim ersten Gliede abbrechen k\"onnen. Dieser Fall ist --- wenn es sich nicht gerade um die innere Struktur des Atoms, oder Moleküls handelt --- bei der makroskopischen physikalischen Anwendung wohl im allgemeinen als realisiert anzusehen. Wir kommen dann zur gewohnten (physikalischen) L\"osung der Potentialgleichung.

   Es fragt sich nun, wie k\"onnen die Koeffizienten dieser Reihe, ebenso wie auch der r\"aumliche Verlauf der Singularit\"at, welche ja in der Natur --- wenigstens in der anorganischen --- nichts weniger als willkürlich, vielmehr in jedem Augenblicke durch \"aussere Einwirkungen eindeutig determiniert sind, bestimmt werden. Fragen solchen Charakters werden in der Physik in klassischer Weise durch Aufstellung von Variationsprinzipen beantwortet und insbesondere ist es das H{\scriptsize AMILTON}sche Prinzip, dem die universellste Bedeutung zuzukommen scheint. Wir wollen jetzt seine unserer Auffassung entsprechende Formulierung versuchen.

   Das H{\scriptsize AMILTON}sche Prinzip sagt aus, dass das sogenannte Wirkungsintegral:
\begin{equation}
W = \int_{\tau_1}^{\tau_2} (T - \Psi \, ) \, d \tau
\end{equation}
bei gegebenem Anfangs- und Endzustand des Systems für das tats\"achlich eintretende Geschehen das kleinstm\"ogliche wird. Dabei ist $T$ die gesamte kinetische, $\Psi$ die potentielle Energie des Systems. Im elektromagnetischen Felde ist die kinetische Energie durch die elektrische, die potentielle durch die magnetische Energie vertreten. Um die ganze Energie des Feldes zu erhalten, muss man über den ganzen elektromagnetischen Raum integrieren, da ja die Energie r\"aumlich verbreitet ist. Das ergibt eine dreidimensionale Integration nach den r\"aumlichen Koordinaten und nun kommt noch im H{\scriptsize AMILTON}schen Prinzip eine Integration nach der Zeit hinzu. Wir sehen also sofort, wie sich die Sache im Sinne des Relativit\"atsprinzips gestalten wird.  {\it Wir werden eine Integration über den gesamten vierdimensionalen Raum auszufüren haben} und den Grenzwert dieses Integrales suchen müssen. Der Integrand soll hierbei bloss vom Werte der elektromagnetischen Grundfunktion oder ``Feldst\"arke'' $\mathcal{F}$ im betreffenden Punkte abh\"angen und, als rein skalare Gr\"osse, einer L{\scriptsize ORENTZ}-Transformation gegenüber invariant bleiben. Daraus folgt schon im vorhinein, dass wir notwendigerweise die absolute Gr\"osse von $\mathcal{F}$, bzw., ihr Quadrat:
\begin{equation}
\mathcal{F} \, \CON{\mathcal{F}} = X^2 + Y^2 + Z^2 + T^2
\end{equation}
in Betracht ziehen werden müssen, als einzige Invariante eines Vektors. Dasselbe gilt auch, wenn die Feldst\"arke nicht der Vektortransformation gehorcht, sondern noch einer $\mathcal{Q}$-Transformation unterworfen wird. Denn es ist auch:
\begin{equation}
\mathcal{F} \, q_0 \, \CON{\mathcal{F} \, q_0} = \mathcal{F} \, \CON{\mathcal{F}}
\end{equation}
Man kann somit schon aus blossen Dimensionsbetrachtungen mit Zuziehung des Relativit\"atsprinzipes die Form des Wirkungsintegrales ansetzen, n\"amlich als:
\begin{equation}
\int \mathcal{F} \, \CON{\mathcal{F}} \, dx \, dy \, dz \, d \tau
\end{equation}
Dieser Ausdruck harmonisiert aber in wunderbarer Weise mit der zu Anfang hingeschriebenen Form des H{\scriptsize AMILTON}schen Prinzipes. Es ist n\"amlich:
\begin{equation}
\mathcal{F} \, \CON{\mathcal{F}} = (\mathcal{H} +i\mathcal{E}) (\CON{\mathcal{H}} +i\CON{\mathcal{E}}) = \mathcal{H} \, \CON{\mathcal{H}} -\mathcal{E} \, \CON{\mathcal{E}} +i ( \mathcal{E}  \, \CON{\mathcal{H}} + \mathcal{H} \, \CON{\mathcal{E}} )
\end{equation}
Der reelle Teil ist somit gleich dem Quadrat der magnetischen Feldst\"arke vermindert um das Quadrat der elektrischen Feldst\"arke. Führen wir zuerst die Integration nach den r\"aumlichen Koordinaten bei konstanter Zeit aus, so erhalten wir tats\"achlich die Differenz der im ganzen Felde enthaltenen magnetischen und elektrischen Energie, (auf einem vom benutzten Massystem abh\"angigen konstanten Faktor kommt es hierbei nicht an), im Sinne der M{\scriptsize AXWELL}schen Theorie und der gew\"ohnlichen dreidimensionalen Anschauung entsprechend. Bei einer solchen Scheidung der Variabelen würde aber die vierdimensionale Welt als in einen Kreiszylinder mit unendlich grossen Abmessungen eingeschrieben vorgestellt sein, was dem Geiste des Relativit\"atsprinzips widerspricht. Die gegebene Anfangs- und Endlage der Welt w\"are die untere, bzw., obere Basis dieses Zylinders. Wir müssen eine relativtheoretisch zul\"assige Begrenzung des Weltalls suchen und die L\"osung dieser Frage ist für uns selbst dann unausweichlich notwendig, wenn diese Grenzen im Unendlichen liegen (denn auch dann muss die Forderung: ``Integration über den ganzen Raum'' durch einen entsprechenden Grenzübergang vollzogen werden.)\footnote{Auch eine Kugel mit unendlich grossem Radius entspricht nicht, da ja die vierte Variabele nicht die Zeit : $\tau$, sondern $i \tau $ ist.}

   Besch\"aftigen wir uns vor allem mit dem vorgeschriebenen Anfangs- und Endzustand der Welt, zwischen welchen die Variation zu vollstrecken ist. Es w\"are sehr unwahrscheinlich, dass jeder Weltlinie irgendein bestimmter Punkt als Anfang und einer als Ende vorgeschrieben w\"are und diese Grenzpunkte auf irgendwelche Art in der Welt zerstreut w\"aren. Ich glaube, hier dr\"angt sich mit einer fast intuitiven Gewissheit der Gedanke auf, dass ein $\emph{einziger}$ Punkt als Anfang und ein einziger Punkt als Ende der Welt anzunehmen ist. Das heisst, dass ein einziger Punkt den Austrittspunkt aller Weltlinien bildet, ebenso wie es einen Punkt gibt, wo alle Weltlinien endigen müssen. Wir haben es doch nur mit einer einzigen Funktion $\mathcal{F}$ zu tun, von welcher die verschiedenen Singularit\"aten nur je einen Teil ausmachen, w\"ahrend sie erst in ihrer Gesamtheit die ganze Funktion ergeben. Diese Einheit wird nun dadurch zur Tatsache gemacht, dass sie alle einen gemeinsamen Anfangs- und Endpunkt besitzen und so gewissermassen eine einzige geschlossene Linie bilden.

\begin{figure}
\begin{center}
\resizebox{16cm}{!}{ \includegraphics{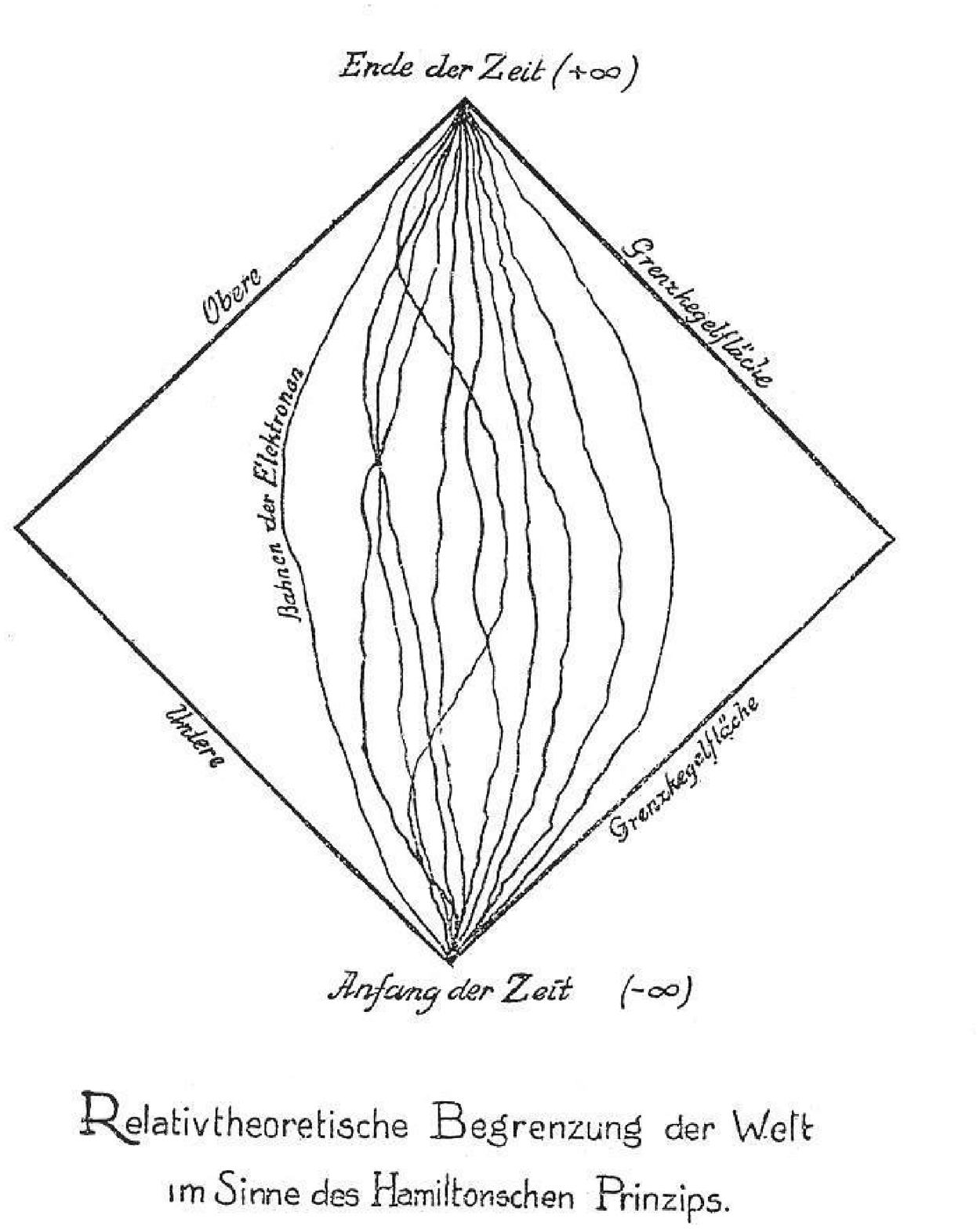}}
\end{center}
\caption{Relativtheoretische Begrenzung der Welt}
\end{figure}

   Bei dieser Annahme wird auch die Frage nach der Begrenzung der Welt eindeutig beantwortet. Da n\"amlich keine \"{U}berlichtgeschwindigkeit eintreten kann, müssen alle Weltlinien notwendigerweise innerhalb jener Kugel bleiben, welche sich vom Anfangspunkte aus mit Lichtgeschwindigkeit ausbreitet, ebenso, wie sie innerhalb einer Kugel bleiben müssen, welche sich zum Endpunkte hin mit Lichtgeschwindigkeit kontrahiert (damit sie letzteren erreichen k\"onnen). In der vierdimensionalen Abbildung f\"allt also die ``m\"ogliche'' Welt einerseits ins Innere des Kegels:
\begin{equation}
\tau = \tau_1 + \sqrt{x^2 + y^2 + z^2}
\end{equation}
andererseits ins Innere des Kegels:
\begin{equation}
\tau = \tau_2 - \sqrt{x^2 + y^2 + z^2}
\end{equation}
Hierbei ist das Achsenkreuz so gelegt, dass die Koordinaten der beiden Grenzpunkte $(0, 0, 0, \tau _1 \, )$ bzw. $(0, 0, 0, \tau _2 \, )$ seien. Diese beiden Kegel (s.\ Abbildung 6.1, welche r\"aumlich vorzustellen ist,) schliessen einen bestimmten Raum ein, dessen Grenzen durch keinerlei Mittel erreicht oder gar überschritten werden k\"onnen und auf diesen Raum ist die vierdimensionale Integration im H{\scriptsize AMILTON}schen Prinzip auszustrecken, da hier nur der physikalisch m\"ogliche Raum in Betracht kommen kann. Es verhindert aber nichts, die Grenzpunkte ins Unendliche hinauszurücken, d.h., die Welt in zeitlicher, wie r\"aumlicher Hinsicht ohne endliche Grenzen vorzustellen. Die beiden Kegelspitzen müssen dann als Limes aufgefasst werden, denen die Elektronenbahnen mit der Zeit unendlich nahe kommen, ohne aber den einen in der Vergangenheit je erreicht zu haben, oder den anderen in der Zukunft je zu erreichen. Die Grenzkegelfl\"achen rücken dann natürlich auch ins Unendliche hinaus. Auf alle F\"alle wird hier eine Richtung gewissermassen als Achse der Welt den anderen gegenüber ausgezeichnet, n\"amlich jene Gerade, welche die beiden Grenzkegelspitzen verbindet. Somit scheint's also, als ob das Relativit\"atsprinzip a priori gefordert, a posteriori wieder aufgehoben würde. Es liegt jedoch im Wesen der Variationsprinzipe, dass die Lage der Grenzen für den Verlauf der Funktion irrelevant ist. (Eine geod\"atische Linie zwischen zwei bestimmten Punkten z.B., bleibt auch in Bezug auf beliebige zwei andere ihrer Punkte geod\"atisch.)

   Das H{\scriptsize AMILTON}sche Integral ist seiner Dimension nach eine sognannte Wir\-kungs\-gr\"osse. Wir k\"onnen also den Integrand, das Quadrat der Feldintensit\"at, als die in der Volumeinheit enthaltene Wirkungsgr\"osse bezeichnen und damit dem H{\scriptsize AMILTON}schen Prinzip die Fassung geben: \emph{die gesamte in der Welt enthaltene Wirkungsgr\"osse muss zwischen allen überhaupt m\"oglichen Werten ein Extremwert sein.} Es soll schon hier bemerkt werden, dass zu dieser Forderung noch Nebenbedingungen in der Gestalt von Vorschriften für die Grenzkegelfl\"achen hinzukommen k\"onnen. Eine solche Bedingung ist zum Beispiel, dass auf der unteren Kegelfl\"ache die Feldst\"arke durchwegs gleich Null sein soll. Die Konsequenz dieser Forderung ist, dass die Elektronen nur in die Zukunft, nicht aber in die Vergangenheit zurück ausstrahlen k\"onnen (--- also keine sich kontrahierenden Kugelwellen m\"oglich sind).

   Wir wollen uns nunmehr zur expliziten Ausrechnung des H{\scriptsize AMILTON}schen Integrals zuwenden. Es zeigt sich, dass die Integration über einen regul\"aren Raum durch ein Oberfl\"achenintegral ersetzt werden kann. Denn aus den beiden Gleichungen:
\begin{equation}
\frac{\partial  \CON{\mathcal{F}}}{\partial x } \, 1_x + \frac{\partial  \CON{\mathcal{F}}}{\partial y } \, 1_y + \frac{\partial  \CON{\mathcal{F}}}{\partial z } \, 1_z + \frac{\partial  \CON{\mathcal{F}}}{\partial t } = 0 
\end{equation}
\begin{equation}
\mathcal{F} = 1_x \, \frac{\partial \, \CON{\Phi}}{\partial x } + 1_y  \, \frac{\partial \, \CON{\Phi}}{\partial y } + 1_z \, \frac{\partial   \, \CON{\Phi}}{\partial z } + \frac{\partial \, \CON{\Phi}}{\partial t } 
\end{equation}
folgt:
\begin{equation}
\frac{\partial  ~}{\partial x } (\CON{\mathcal{F}} \, 1_x \, \CON{\Phi }) + \frac{\partial  ~}{\partial y } (\CON{\mathcal{F}} \, 1_y \, \CON{\Phi }) + \frac{\partial  ~}{\partial z } (\CON{\mathcal{F}} \, 1_z \, \CON{\Phi }) + \frac{\partial  ~}{\partial t } (\CON{\mathcal{F}} \, \CON{\Phi }) = \CON{\mathcal{F}} \, \mathcal{F}
\end{equation}
Auf der linken Seite k\"onnen wir unmittelbar den G{\scriptsize AUSS}schen Satz anwenden und erhalten so, indem wir mit $dv$ das vierdimensionale Volumelement, mit $df$ das Fl\"achenelement der Begrenzungsfl\"ache bezeichnen, die Beziehung:
\begin{equation}
\int \mathcal{F} \, \CON{\mathcal{F}} \, dv = \int \CON{\mathcal{F}} \, n \, \CON{\Phi} \, df
\end{equation}
Damit identisch ist auch:
\begin{equation}
\int \Phi \, \CON{n} \, \mathcal{F} \, df
\end{equation}
$n$ bedeutet wieder die nach aussen weisende Fl\"achennormale mit der L\"ange Eins.

   Bei der Integration über den ganzen Raum sind die Singularit\"aten einer besonderen Behandlung bedürftig. Sie müssen durch geschlossene Fl\"achen umgeben vom Integrationsgebiet ausgeschlossen, diese Fl\"achen immer inniger angeschmie\-gt und im Limes mit den Singularit\"aten vereinigt werden. Das r\"aumliche Integrationsgebiet wird also einerseits durch diese die Singularit\"aten umgebenden Fl\"achen, andererseits durch die Grenzkegelfl\"achen begrenzt, auf diese muss also auch das statt dem ursprünglichen Raumintegrals nehmbare Oberfl\"achenintegral ausgestreckt werden. Die \"aussere Grenze nehmen wir im Unendlichen an und wollen uns vorl\"aufig mit der Gesamtheit der Singularit\"aten, also den Bahnen der Elektronen besch\"aftigen.

   \"{U}ber ihren speziellen Aufbau und Struktur k\"onnen wir allerdings nichts Bestimmtes aussagen, wissen wir doch nicht einmal, ob diese Bahnen einfache Linien (punktf\"ormiges Elektron) oder kompliziertere Gebilde vorstellen sollen. Glücklicherweise kann aber ein Teil des Integrals, und zwar --- wie es scheint ---, für viele F\"alle der ausschlaggebende, ausgerechnet werden, ohne auf irgendwelche Einzelheiten einzugehen. Wir k\"onnen in der Umgebung einer Singularit\"at die Feldst\"arke in zwei Teile zerlegen: in eine \"aussere: $\mathcal{F}_e $ welche von den \"ausseren Singularit\"aten (den übrigen Elektronen) herrührt und eine innere: $\mathcal{F}_i $, welche das ``Eigenfeld'' des in Frage stehenden Elektrons darstellt. Ebenso ist es mit dem Potential $\Phi$. Wir haben also den folgenden Ausdruck zu berechnen:
\begin{equation}
\int (\,\Phi_e + \Phi_i \, ) \, \CON{n} \, ( \, \mathcal{F}_e + \mathcal{F}_i \, ) \, df
\end{equation}
Das ganze Integral ist somit in vier Summen zerlegbar, die wir einzeln in Betracht ziehen wollen. Die Integration ist über eine r\"ohrenartige, die ganze Singularit\"at umschliessende Fl\"ache auszubreiten. Wir zerlegen sie in lauter Elementarzylinder in der Art der
\begin{figure}
\begin{center}
\resizebox{6cm}{!}{ \includegraphics{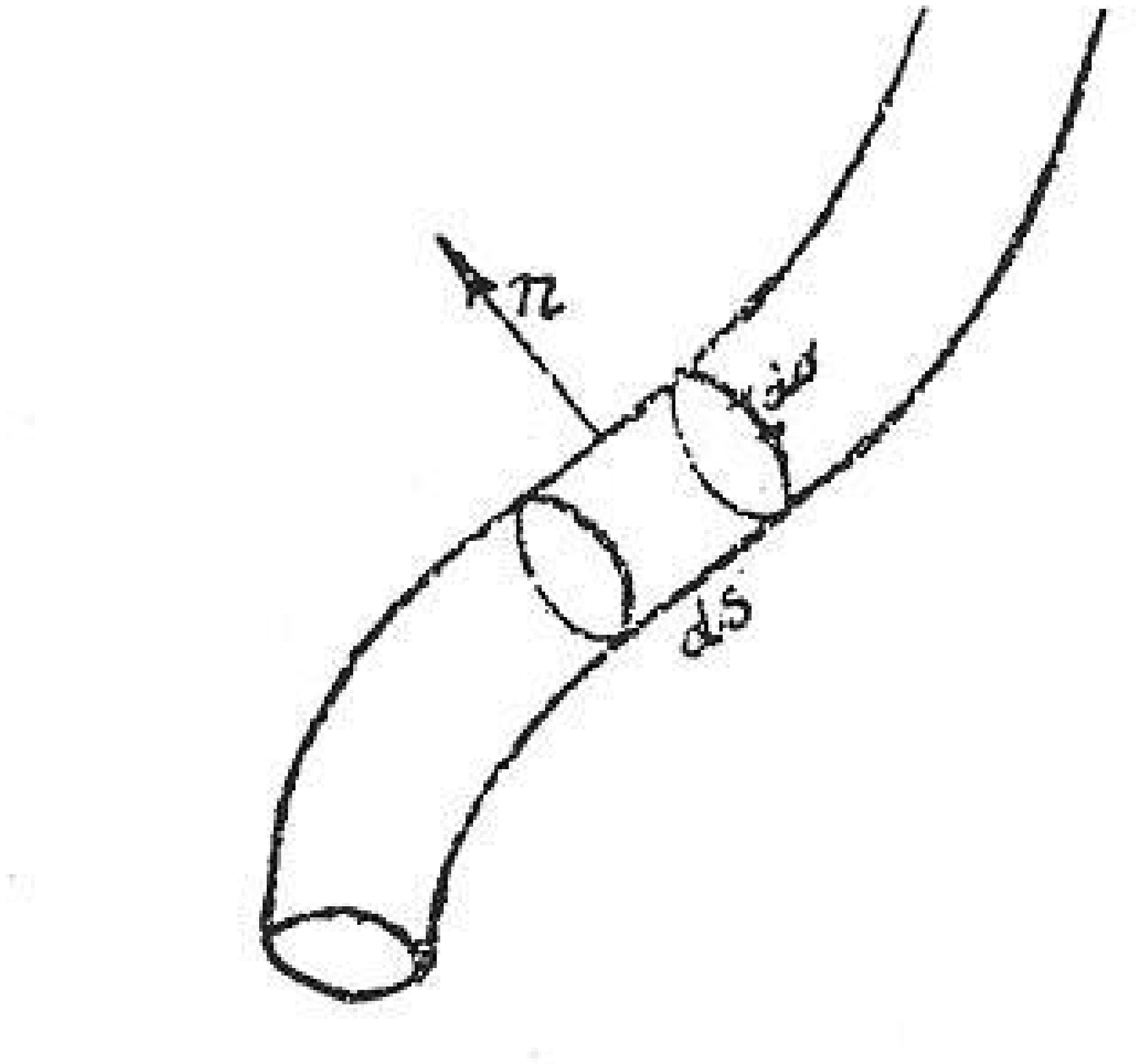}}
\caption{R\"ohrenartige Integrationsfl\"ache}
\end{center}
\end{figure}
\noindent Figur 6.2, ihre H\"ohe soll mit $ds$ bezeichnet sein. Im Falle des punktf\"ormigen Elektrons ist $ds$ das Linienelement der Bahn. Die Integration führen wir nun derart aus, dass wir zuerst l\"angs der Basislinie des Zylinders --- dem Querschnitt der R\"ohrenfl\"ache --- herumintegrieren und dann nach der L\"ange. Bei der ersten Integration wird die \"ausser Feldst\"arke und ebenso, das \"aussere Potential als homogen angesehen werden dürfen, wenn, wie dies gew\"ohnlich der Fall ist, die eventuellen Dimensionen des Elektrons gegenüber den gegenseitigen Entfernungen zu vernachl\"assigen ist. Wir k\"onnen also $\mathcal{F}_e $ und $\Phi_e $ vor das Integrationszeichen herausheben und erhalten so die beiden Ausdrücke:
\begin{equation}
\int \Phi_e \Big( \int \CON{n} \, \mathcal{F}_i \, d \sigma \Big) \, ds
\end{equation}
und:
\begin{equation}
\int \CON{\mathcal{F}}_e \, \Big( \int n \, \CON{\Phi _i} \, d \sigma \Big) \, ds
\end{equation}
Wir setzen:
\begin{equation}
                 L = \lim \int \CON{\mathcal{F}_i} \, n \, d\sigma
\end{equation}
\begin{equation}
                  \Lambda = \lim \int n \, \CON{\Phi _i} \, d\sigma 
\end{equation}
und schliesslich:
\begin{equation}
              H = \lim \int \Phi_i \, \CON{n} \, \mathcal{F}_i \, d\sigma
\end{equation}
$d\sigma$ ist das Element einer Fl\"ache im dreidimensionalen Raume, welche das Elektron umschliesst. Zu jeden Punkte der Bahnlinie geh\"ort ein gewisser Wert von $L$, $\Lambda $ und $H$. Das in Betracht gezogene Elektron steuert also zum Wirkungsintegral mit den drei Integralen:
\begin{equation}
\int \Phi_e \, \CON{L} \, ds \, + \int \CON{\mathcal{F}_e } \, \Lambda \, ds \, + \int \negthinspace H \, ds 
\end{equation}
bei. Der vierte Teil:
\begin{equation}
          \lim \int \Phi_e \, \CON{n} \, \mathcal{F}_e \, df
\end{equation}
ist n\"amlich gleich Null.

	Die Wirkungsgr\"osse des gesamten Raumes erhalten wir somit durch drei auf die Quellen der Feldst\"arke verteilte Summen in der Form:
\begin{equation}
\sum \int \Phi_e \, \CON{L} \, ds + \sum \int \CON{\mathcal{F}_e } \, \Lambda \, ds + \sum \int H \, ds 
\end{equation}
und ausserdem kommt noch das über die \"aussere Grenzfl\"ache erstreckte Integral:
\begin{equation}
              \int \Phi \, \CON{n} \, \mathcal{F} \, df 
\end{equation}
hinzu.

   Die beiden Limes: $L$ und $\Lambda $ existieren tats\"achlich --- wie es sich zeigen l\"asst --- im allgemeinen, und zwar kommt für $L$ --- wir wollen sie die ``Belegung'' bezeichnen --- in der unendlichen Reihenentwicklung für das Potential nur der Koeffizient von $1/r$, für $\Lambda $ der Koeffizient von $1/r^2$ in Betracht. Brechen wir zum Beispiel die Reihe gleich beim ersten Gliede ab, ist also:
\begin{equation}
\Phi =\frac{\varphi(\tau') \sqrt{ 1 - v^2 }}{ r (1 + \dot{r})}
\end{equation}
so ist die Belegung:
\begin{equation}
L= 4 \pi \varphi (\tau)
\end{equation}
w\"ahrend $\Lambda = 0$ ist. Für $\tau$ ist die Zeitkoordinate des betreffenden Punktes der Weltlinie einzusetzen.

   Anders steht die Sache mit dem Grenzwerte $H$, den wir als ``H{\scriptsize AMILTON}sche Funktion des Elektrons'' bezeichnen wollen. Dieser Grenzwert braucht im allgemeinen nicht zu existieren, wenigstens nicht als endliche Gr\"osse. Schon in dem eben betrachteten Beispiel für das Potential $\Phi$ wird der Wert von $H$ unendlich gross. Ohne die allgemeinen Bedingungen für das Endlichbleiben zu untersuchen, m\"ochte ich im folgenden Kapitel ein Elektonenmodell er\"orten, bei welchem die H{\scriptsize AMILTON}sche Funktion Null als Grenzwert besitzt.

\chapter{Das Kreiselektron.}

Das Potential:
\begin{equation}
\Phi = ~\frac{1}{\sqrt{ (x - \xi)^2 + (y - \eta)^2 + (x - \zeta)^2 }}
\end{equation}
erzeugt ein zeitlich station\"ares Feld mit dem Punkte $( \, \xi, \eta, \zeta \, )$ als Singularit\"at. Da wir grunds\"atzlich Potential und Feldst\"arke als komplexe Gr\"ossen eingeführt haben, entspricht es dem Sinne unserer Anschauungen, die Konstanten: $( \, \xi, \eta, \zeta \, )$ als komplexe Zahlen anzusehen. Durch Koordinatenverschiebung und Drehung erreichen wir, dass zwei der Konstanten, z.B., $\xi $ und $\eta $, gleich Null und die dritte rein imagin\"ar angesetzt werden kann:
\begin{equation}
\zeta = -i \varrho
\end{equation}
so das wir haben:
\begin{equation}
\Phi = ~\frac{1}{\sqrt{(x^2 + y^2 + ( z + i \varrho )^2 }}
\end{equation}
Der Nenner wird hier nicht nur für einen einzigen Punkt, sondern l\"angs einer ganzen Kreislinie null, n\"amlich für:
\begin{gather}
\left.
\begin{array}{c}
z   = 0 \\
\text{und} \hskip 4 cm   \\
x^2 + y^2   = \varrho^2
\end{array}
\quad \right\}
\end{gather}
Aber nicht nur diese Kreislinie ist für die Funktion singul\"ar, sondern die ganze durch sie umschlossene Kreisfl\"ache. Bei Durchschreitung dieser Fl\"ache wechselt n\"amlich die Quadraturwurzel ihr Vorzeichen.

   Die Ausrechnung der H{\scriptsize AMILTON}schen Funktion kann in zwei Schritten vorge\-nommen werden. Wir integrieren l\"angs der
\begin{figure}
\begin{center}
\resizebox{8cm}{!}{ \includegraphics{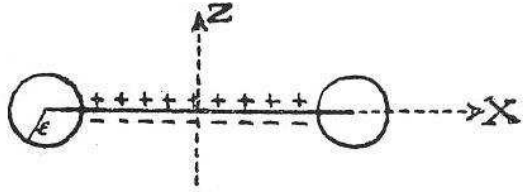}}
\end{center}
\caption{Kreiselektron herumgeschlagene Ringfl\"ache}
\end{figure}
Kreisscheibe oben und unten herum bis zum Radius $\varrho - \varepsilon$, und dann l\"angs einer um den Kreis herumgeschlagene Ringfl\"ache mit dem Radius $\epsilon $ und gehen dann zur Grenze $\varepsilon = 0$ über. Da aber an der Kreisscheibe sowohl $\Phi$, wie auch $\mathcal{F}$ das Vorzeichen wechseln, verh\"alt sich ihr Produkt regul\"ar und darum wird das Integral über die Kreischeibe gleich Null. Für die Ringfl\"ache genügt es wegen der allseitigen Symmetrie um die $Z$-Achse herum die Rechnung in der Projektion $y=0$ auszuführen, um dann mit $2\pi \varrho$ zu multiplizieren. Wir setzen:
\begin{gather}
\left.
\begin{array}{c}
      x = \varrho + \varepsilon \cos \varphi  \\
      z = \varepsilon \sin \varphi
\end{array}
\quad \right\}
\end{gather}
und erhalten als Resultat:
\begin{equation}
H=\frac{2 \pi \varrho}{\varepsilon} \int_0^{2 \pi} \frac{\varepsilon + \varrho (\cos \varphi + i \sin \varphi)}{ [\varepsilon + 2 \varrho (\cos \varphi + i \sin \varphi)]^2} \, d \varphi
\end{equation}
Man überzeugt sich leicht, dass dieses Integral den Wert Null hat (und zwar von $\varepsilon$ unabh\"angig.)

   Das Kreiselektron gibt uns ein einfaches Beispiel, dass die H{\scriptsize AMILTON}sche Funktion des Elektrons nicht notwendig unendlich werden muss. Im station\"aren Felde bedeutet die H{\scriptsize AMILTON}sche Funktion die Differenz der elektrischen und magnetischen Energie. Beim punktf\"ormigen Elektronen, welches ein rein elektrisches Feld erzeugt, wird die Feldenergie unendlich gross. Auch beim Kreiselektron wird die elektrische Energie unendlich, es tritt jedoch auch ein magnetisches Feld auf, dessen ebenfalls unendlich werdende Energie die elektrische gerade kompensiert, so dass die Differenz der beiden dem Grenzwerte Null zustrebt. Die magnetischen Kr\"afte werden nur in der N\"ahe des Elektrons bemerkbar. In grosser Entfernung, welcher gegenüber der Radius $\varrho$ des Elektrons zu vernachl\"assigen ist, unterscheidet sich das Feld des Kreiselektrons von dem des Punktelektrons nur unendlich wenig.

\chapter{Dynamik des Elektrons im Gravitationsfeld
                und im\\ elektromagnetischen Felde.}

Das Oberfl\"achenintegral, in welches das Wirkungsintegral des Raumes umgewandelt wurde, ist ausser über die Singularit\"aten noch über die \"aussere Begrenzungsfl\"ache des Raumes auszubreiten. Der Beitrag dieser Fl\"ache darf keineswegs einfach Null gesetzt werden, selbst wenn die Grenzen im Unendlichen liegen. Es ist vielmehr wahrscheinlich, dass der Grenzfl\"ache eine charakteristische Rolle zukommen wird. Wir haben wohl die Grenzkegelfl\"achen des m\"oglichen Raumes relativtheoretisch gerechtfertigt, wenn aber der funktionentheoretische Standpunkt richtig ist, so müssen diese Grenzen auch funktionentheoretische Bedeutung haben. Dies ist der Fall, wenn s\"amtliche Punkte der Grenzfl\"ache singul\"ar sind, dann ist die Funktion nicht über das durch sie eingeschlossene Gebiet hinaus fortsetzbar. ``Solche'' Funktionen mit natürlichen Grenzen geh\"oren hier nicht zu den Ausnahmen, sie bilden im Gegenteil gewissermassen den natürlichen \"{U}bergang vom reellen Raum (mit der Variabelen: $t$) zum imagin\"aren (mit der Variabelen: $i \tau$). Einer punktf\"ormigen Singularit\"at im reellen vierdimensionalen Raume, dessen Grundtypus die Funktion:
\begin{equation}
\Phi = \frac{1}{r^2 + t^2}
\end{equation}
bildet, entspricht n\"amlich im imagin\"aren Raum, wie die Funktion:
\begin{equation}
\Phi = \frac{1}{r^2 - \tau^2}
\end{equation}
zeigt, eine Kegelfl\"ache als singul\"are Fl\"ache, und zwar die beiden Kegel: $r=t$, oder: $r=-t$. Dasselbe ist der Fall bei einer unendlichen Reihe, welche durch Differentiation dieser Funktion nach den einzelnen Variabelen und Summierung entsteht. Setzen wir für beide Grenzpunkte der Welt je eine solche Reihe an und addieren diese zu den von den inneren Singularit\"aten entspringenden Reihen, so haben wir schon eine Funktion vor uns, deren natürliche Grenze tats\"achlich durch die beiden Grenzkegelfl\"achen ausgesteckt wird. Diese Entwicklungen treten aber schon zu sehr aus dem gewohnten Rahmen der physikalischen Spekulation heraus, als dass ihre n\"ahere Er\"orterung hier bei einem derart embryonalen Stadium der Theorie am Platze w\"are. Ich m\"ochte darum diese Frage übergehen und vorl\"aufig die Grenzfl\"achen unberücksichtigt lassen.

   Hingegen m\"ochte ich noch die Anwendbarkeit der entwickelten Anschauungen an einem ganz speziellen Beispiel zeigen, n\"amlich durch Ableitung der Bewegungsgleichungen für ein freies Elektron, welches sich in einem Gravitationsfeld, oder ausserdem noch in einem elektromagnetischen Felde befindet. Man kann bei der physikalischen Anwendung der Variationsprinzip den in der Mechanik üblichen glücklichen Kunstgriff benutzen, einen verborgenen Mechanismus durch empirisch gefundene Bedingungsgleichungen zu ersetzen, d.h., das unbekannte System durch eine postulativ angenommene Verminderung der Anzahl der Freiheitsgrade gewissermassen auszuschalten. Wir werden in \"ahnlichem Sinne die allgemeine Aufgabe sehr betr\"achtlich einschr\"anken. Erstens wollen wir in der unendlichen Reihenentwicklung des Potentials nur das erste Glied berücksichtigen, zweitens soll ausserdem die Belegung empirisch postuliert werden, so das als einzige Freiheit der Verlauf der Bahnlinie des als punktf\"ormig betrachteten Elektrons übrigbleibt. Das Elektron ist also hier ein System mit drei Freiheitsgraden, es sollen seine r\"aumlichen Koordinaten als Funktionen der Zeit bestimmt werden.

   In den drei Teilintegralen, welche das ganze Wirkungsintegral ausmachen, ist bei der vorausgesetzten Einschr\"ankung $\Lambda =0$ zu setzen. Somit f\"allt dieser Teil weg. Weiterhin sollen die H{\scriptsize AMILTON}schen Funktionen der Elektronen gleich Null, oder wenigstens den in Betracht kommenden Gr\"ossen gegenüber als verschwindend klein angesehen werden. Es bleibt somit nur die einzige Summe übrig:
\begin{equation}
 \sum \int \Phi_e \, \CON{L} \, ds
\end{equation}
wobei:
\begin{equation}
\qquad \qquad \Phi = \sum \frac{L \, \sqrt{1-v^2}}{4 \pi r \, (1+\frac{\partial  r}{\partial \tau ' } )} \qquad  \qquad ( \, \tau ' = \tau - r \, )
\end{equation}
ist.

   Es fragt sich vor allen Dingen in welcher Weise im allgemeinen die Integration vorzunehmen ist, um die gewohnten E{\scriptsize ULER}schen Gleichungen als L\"osung ansetzen zu k\"onnen. W\"ahlen wir ein bestimmtes Elektron aus. Um mit ihm die Variation auszuführen, genügt es, nur jenen Teil des Wirkungsintegrals zu kennen, in welchem die Daten dieses Elektrons vorkommen. Vor allem muss also über seine eigene Bahn integriert werden. Ausserdem fliesst es aber auch bei allen übrigen Elektronen in das \"aussere Potential $\Phi_e$ ein, so dass zu gleicher Zeit auch alle anderen Bahnen berücksichtigt werden müssen. Die Punkte der Bahnlinien aller übrigen Elektronen sollen nun folgendermassen zu den Punkten der eigenen Bahn zugeordnet werden. Zu jedem Punkte soll von den übrigen Elektronen immer jener Punkt der Bahnlinie
\begin{figure}
\begin{center}
\resizebox{8cm}{!}{ \includegraphics{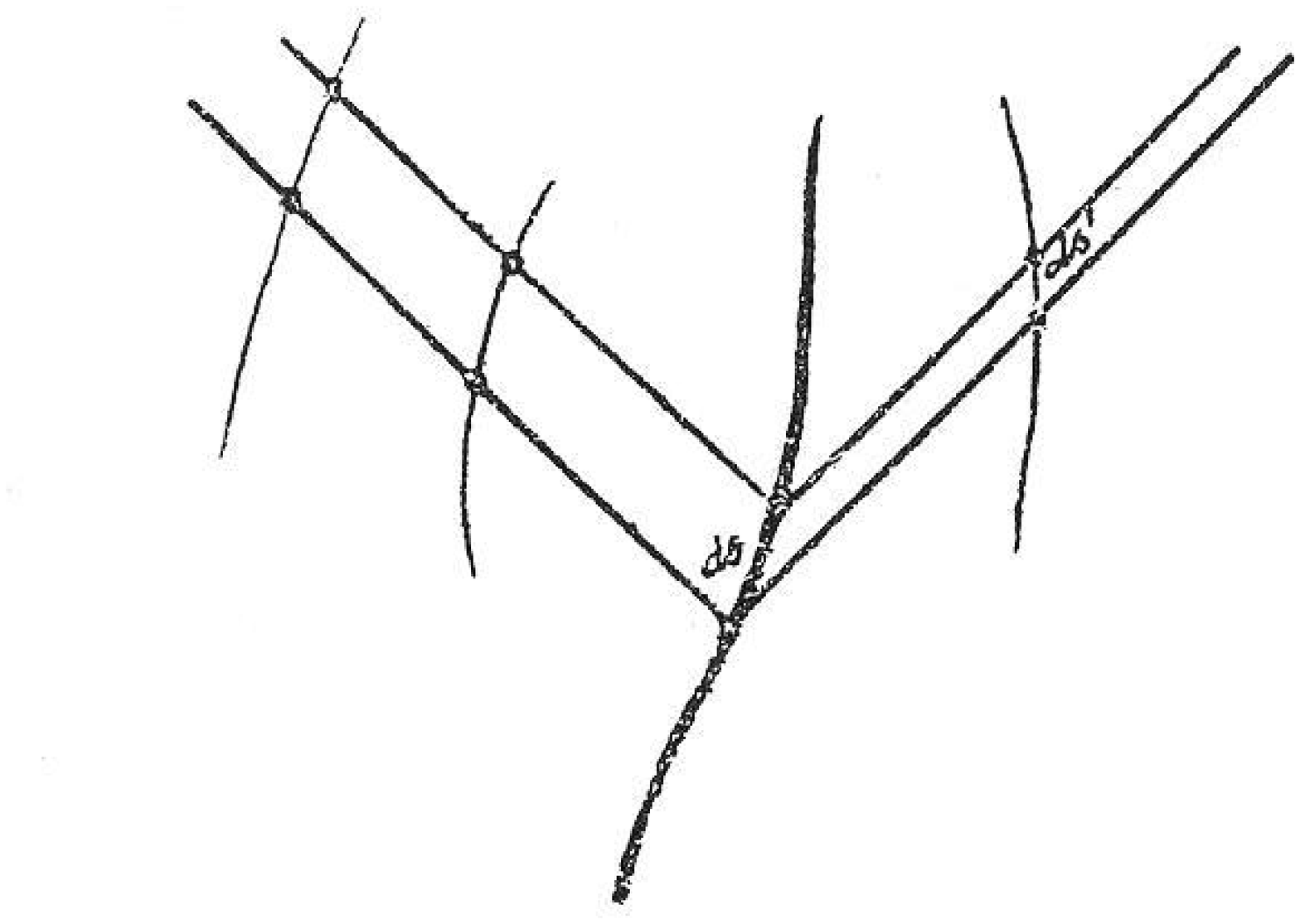}}
\caption{Elektronenbahnlinien}
\end{center}
\end{figure}
hinzugeh\"oren, welchen die vom betreffenden Aufpunkte aus mit der Einheit der Geschwindigkeit forteilende Welle gerade erreicht. Die Zeitpunkte der koordinierten Punkte h\"angen also mit der Zeit $\tau$ des Aufpunktes durch die Gleichung:
\begin{equation}
\tau ' = \tau + r
\end{equation}
zusammen, wobei unter $r$ die Entfernung des Aufpunktes zur Zeit $\tau$ mit dem koordinierten Punkte zur Zeit $\tau'$ zu nehmen ist. $r$ h\"angt somit sowohl von $\tau$, wie von $\tau'$ ab, denn es ist:
\begin{equation}
r^2 = [ \, \xi ' (\tau ') - \xi (\tau) \, ]^2 +[ \, \eta ' (\tau ') - \eta (\tau) \, ]^2 +[ \, \zeta ' (\tau ') - \zeta (\tau) \, ]^2
\end{equation}
und weiterhin:
\begin{equation}
d \tau ' = d\tau + \frac{\partial  r}{\partial \tau ' } \, d\tau ' + \frac{\partial  r}{\partial \tau} \, d\tau
\end{equation}
also:
\begin{equation}
d \tau ' = \frac{1 + \frac{\partial  r}{\partial \tau}}{1 - \frac{\partial  r}{\partial \tau ' }} d\tau 
\end{equation}
   Wir fassen jetzt alle Glieder zusammen, welche mit dem Differential der Bahnl\"an\-ge: $ds$ multipliziert sind und von der Lage des Aufpunktes: $( \, \xi, \eta, \zeta \, )$ abh\"angen. Sie k\"onnen in der Form geschrieben werden:
\begin{equation}
( \, \Phi_e + \Phi '_e \, ) \, \CON{L}
\end{equation}
wobei:
\begin{equation}
{~} \qquad \qquad \Phi '_e  = \sum  \frac{1}{4 \pi} \frac{L' \, \sqrt{1-v'{}^2}}{r(1-\frac{\partial  r}{\partial \tau ' } )}
\qquad ( \, \tau {}' = \tau + \dot{r} \, )
\end{equation}
ist. Dieses Potential stellt dem retardierten Potential $\Phi_e$ ein ihm \"aquivalentes, jedoch aus der Zukunft zurückstrahlendes zur Seite, welches jedoch auf des retardierte Potential zurückführbar ist.

   Zu diesem Zwecke nehmen wir zum Variationsintegral noch ein Integral hinzu, welches in seiner Totalit\"at den Wert Null hat. Es ist das über den Kegel:
\begin{equation}
\tau = \tau_0 - r
\end{equation}
genommene Integral:
\begin{equation}
2 \int \Big[ \, \Xi \, \Big( \, \frac{\partial \, \Xi }{\partial x} \, a
                               +\frac{\partial \, \Xi }{\partial y} \, b
                               +\frac{\partial \, \Xi }{\partial z} \, c
                               -\frac{\partial \, \Xi }{\partial t} \, d
                                    \, \Big) + .~.~.  \, \Big] \, df
\end{equation}
wobei $\tau_0$ der Grenze $+\infty$ zustreben soll, $\Xi, H, Z, \Theta \,$ die Komponenten des gesamt\-en Potentials, und $a , b , c , d$ die Komponenten der Fl\"achennormale bedeuten. W\"ah\-len wir denjenigen  Teil aus, welcher von der $X$-Komponente der Belegung des in Betracht gezogenen Elektrons abh\"angt, so finden wir den Betrag:
\begin{equation}
\int \frac{L_x}{4\pi R} \, \Big[ \int \Big( \, \frac{\partial \, \Xi_e }{\partial x } \, a +\frac{\partial \, \Xi_e }{\partial y} \, b +\frac{\partial \, \Xi_e }{\partial z} \, c -\frac{\partial \, \Xi_e }{\partial t} \, d \, \Big) \, d \sigma \Big] \, ds
\end{equation}
so das zur $X$-Komponente des Potentials $\Phi {}'_e$ noch hinzukommt:
\begin{equation}
A_x = \frac{1}{4\pi R} \int \Big( \frac{\partial  \, \Xi_e }{\partial x } \, a +\frac{\partial \, \Xi_e }{\partial y} \, b +\frac{\partial \, \Xi_e }{\partial z}  \, c +\frac{\partial \, \Xi_e }{\partial \tau} \, d \, \Big) \, d \sigma
\end{equation}
wenn $( \, a , b , c \, )$ die Normale der unendlich fernen Kugelfl\"ache mit dem Radius $R$, $d\sigma$ das Fl\"achenelement ist, und die Werte der Differentialquotienten zur Zeit $\tau + R$ zu nehmen sind. Die Summe: $\Phi_{e\,x}' + A_x $ ist aber nichts anderes, als die $X$-Komponente des \"ausseren Potentials im Punkte: $( \, \xi , \eta , \zeta \, )$. Wir k\"onnen also statt $\Phi_e'$ $\Phi_e$ in die Formel einführen und erhalten als einzige Gr\"osse, welche bei der Variation in Betracht kommt:
\begin{equation}
                      2 \int \CON{\Phi_e} \, L \, ds
\end{equation}

    Wir haben schon einmal erw\"ahnt, dass zum H{\scriptsize AMILTON}schen Prinzip noch als Nebenbedingungen gewisse für die Grenzfl\"achen geltenden Forderungen hinzu\-kommen k\"onnen. Eine solche Nebenbedingung führen wir jetzt ein, indem wir verlangen, das die Gesamtbelegung der Grenzfl\"ache, d.h., dass auf sie bezogene Fl\"achenintegral:
\begin{equation}
\int \CON{\mathcal{F}} \, n \, df
\end{equation}
einen vorgeschriebenen Wert habe. Durch Anwendung der Grundgleichungen in der Umformung mit Hilfe des G{\scriptsize AUSS}schen Satzes folgt, dass diese Grenzbelegung gleich der Summe s\"amtlicher Belegungen der Elektronen ist, n\"amlich:
\begin{equation}
\sum \int L \, ds
\end{equation}
Nach der Methode der L{\scriptsize AGRANGE}schen Koeffizienten haben wir die Nebenbedingung mit einer Konstanten multipliziert, zum Variationsintegral zu addieren. Diese Konstante sei in unseren Falle: $-2 \Gamma$. Sie bewirkt, dass das \"aussere Potential um den Wert $\Gamma$ vermindert wird, so dass schliesslich für ein bestimmtes Elektron die zu varierende Gr\"osse --- von einem konstanten Faktor abgesehen und statt der Bahnl\"ange die Zeit als unabh\"angige Variabele eingeführt --- wird:
\begin{equation}
\int_{\tau_1}^{\tau_2} ( \, \Phi_e - \Gamma \, ) \, \CON{L} \, \sqrt{1-v^2}  \, d\tau
\end{equation}
Jetzt kann die E{\scriptsize ULER}sche Gleichung unmittelbar angewendet werden, wenn nur die Belegung: $L$ bekannt ist.

   Wir nehmen nun f\"ur das Gravitationsfeld an, dass s\"amtliche Belegungen nach einer bestimmten, auch zeitlich unver\"anderlichen Richtung weisen (es ist plausibel die ``Weltachse,'' d.h., die beiden Grenzpunkte verbindende Gerade als diese universelle Richtung anzunehmen). Wir verlegen die Zeitachse des Koordinatensytems in diese Richtung. Die Belegung und das Potential haben dann nur zeitliche Komponenten, sind somit als Skalare anzusehen. Auch der Gr\"osse nach nehmen wir die Belegung --- die mechanische ``Masse'' --- als zeitlich unver\"andererlich an. Ihre Bezeichnungen sei $\mu$. Wir haben dann:
\begin{equation}
\int_{\tau_1}^{\tau_2} ( \, \Theta^g_e - \Gamma \, ) \, \mu \, \sqrt{1-\dot{\xi}^2-\dot{\eta}^2-\dot{\zeta}^2} \, d\tau
\end{equation}
Die Konstante $\Gamma$ ist ebenfalls eine skalare Gr\"osse. Ihr Wert ist gegenüber den uns bekannten Graviationspotentialen überwiegend gross, n\"amlich im C.G.S.-System ausgedrückt in Zusammenhang mit der Lichtgeschwindigkeit $c$ und der gew\"ohnlichen Gravitationskonstante $\gamma$:
\begin{equation}
\Gamma = \frac{c^2}{4 \pi \gamma} = 10 \times 10^{27} \, \frac{ \text{gr} }{ \text{cm} }
\end{equation}
w\"ahrend das Potential auf der Sonnenoberfl\"ache blos $3 \times 10^{22}$ ist.

   Wir wenden jetzt auf den Integranden $I$ in Bezug auf die Komponenten der Bahn die E{\scriptsize ULER}sche Gleichung an, z.B., für die $X$-Komponente:
\begin{equation}
\frac{d~}{d \tau } \frac{\partial  I}{\partial \dot{\xi} } - \frac{\partial  I}{\partial \xi } = 0
\end{equation}
Wir erhalten --- indem wir $\Theta_e^g$ neben $\Gamma$ vernachl\"assigen:
\begin{equation}
\Gamma \, \frac{d~}{d \tau } \frac{\mu \, \dot{\xi}}{\sqrt{1-v^2} }=\mu \, \sqrt{1-v^2} \, \frac{\partial \, \Theta^g_e}{\partial x }
\end{equation}
Diese Gleichung enth\"alt mit den entsprechenden für $\eta$ und $\zeta$ erg\"anzt das N{\scriptsize EWTON}\-sche Gravitationsgesetz in seiner relativtheoretischen Fassung --- Verbindung des M{\scriptsize INKOWSKI}schen Kraftvektors mit dem Gradienten des retardierten Potentials --- und zugleich die Aequivalenz von schwerer und tr\"ager Masse.

   Im elektromagnetischen Felde f\"allt die Belegung in die jeweilige Richtung der Bahn, ihre Gr\"osse ist der Vierergeschwindigkeit proportional mit imagin\"aren Proportionalit\"atsfaktor, der Koeffizient von $i$ heisst die Ladung des Elektrons. Hier ist also:
\begin{equation}
L=i e \, \Big( \, \frac{\partial \, \xi}{\partial s } \, 1_x + \frac{\partial \, \eta}{\partial s } \, 1_y + \frac{\partial \, \zeta}{\partial s } \, 1_z + \frac{\partial \, \vartheta}{\partial s } \, \Big)
\end{equation}
oder auch:
\begin{equation}
L= \frac{e}{\sqrt{1-v^2}} \, \big( \, \dot{\xi} \, 1_x + \dot{\eta} \, 1_y + \dot{\zeta} \, 1_z + i \, \big)
\end{equation}
Das elektromagnetische Feld lagert sich über das Gravitationsfeld, d.h., ihre Belegungen müssen addiert werden. Wir wollen bloss den reellen Teil des Variationsintegrales in Betracht ziehen (was mit dem imagin\"aren zu geschehen hat, muss vorl\"aufig dahingestellt bleiben). Dieser ist durch das elektromagnetische Feld mit folgender Gr\"osse zu erg\"anzen:
\begin{equation}
e \, \big( \, \Xi_e \, \dot{\xi} +  H_e \, \dot{\eta} + Z_e \, \dot{\zeta} + \Theta_e \, i \, \big)
\end{equation}
$\Xi, H, Z, \Theta$ bedeuten hier die Komponenten des elektromagnetischen ``Vektorpotentials.''  Die Nebenbedingung f\"allt wegen der Integrabilit\"at der Belegung fort. Wenden wir die E{\scriptsize ULER}sche Gleichung an, so sehen wir, dass die ``\,Tr\"agheitskraft'' des Gravitationsfeldes unver\"andert bleibt, zur bewegenden Kraft kommt jedoch folgender Ausdruck in der $X$-Komponente hinzu:
\begin{equation}
e \, \Big( \, \frac{\partial \, \Xi_e }{\partial x } \, \dot{\xi} +\frac{\partial \, H_e }{\partial x} \, \dot{\eta} +\frac{\partial \, Z_e }{\partial x} \, \dot{\zeta} +\frac{\partial \, \Theta_e }{\partial x} \, i \, \Big) - \frac{d~}{d \tau } (e \, \Xi_e \, )
\end{equation}
Es ist aber:
\begin{equation}
\frac{d~}{d \tau } ( \, e \, \Xi_e \, )=
e \, \Big( \, \frac{\partial \, \Xi_e }{\partial x } \, \dot{\xi} +\frac{\partial \, \Xi_e }{\partial y} \, \dot{\eta} +\frac{\partial \, \Xi_e }{\partial z} \, \dot{\zeta} +\frac{\partial \, \Xi_e }{\partial \tau} \, \Big)
\end{equation}
und das Resultat ist also:
\begin{equation}
e \, \Big( \, i \, \frac{\partial \, \Theta_e }{\partial x } - \frac{\partial \, \Xi_e }{\partial \tau } \, \Big) 
+ e \, \dot{\eta} \, \Big( \, \frac{\partial \, H_e }{\partial x } - \frac{\partial \, \Xi_e }{\partial y } \, \Big) 
+ e \, \dot{\zeta} \, \Big( \, \frac{\partial \, Z_e }{\partial x } - \frac{\partial \, \Xi_e }{\partial z } \, \Big)
\end{equation}
Nun finden wir aber durch Spaltung des reellen vom imagin\"aren Teil folgenden Zusammenhang der hier stehenden Gr\"ossen mit der elektrischen und magnetischen Feldst\"arke:
\begin{gather}
\left.
\begin{array}{c}
i \, \dfrac{\partial \, \Theta }{\partial x    }
   - \dfrac{\partial \, \Xi    }{\partial \tau }  = - \mathcal{E}_x \\
~ \\
~ \, \dfrac{\partial \,  H     }{\partial x    }
   - \dfrac{\partial \, \Xi    }{\partial y    }  = - \mathcal{H}_z \\
~ \\
~ \, \dfrac{\partial \,  Z     }{\partial x    }
   - \dfrac{\partial \, \Xi    }{\partial z    }  = - \mathcal{H}_y   
\end{array}
\quad \right\}
\end{gather}
Wir erhalten somit als bewegende Kraft des elektromagnetischen Feldes genau denselben Ausdruck, wie ihn die Elektronentheorie ansetzt, n\"amlich vektoriell geschrieben:
\begin{equation}
-e \, ( \, \mathcal{E}_e + v \, H_e \, )
\end{equation}

   Der merkwürdige Zeichenwechsel des C{\scriptsize OULOMB}schen Gesetzes gegenüber dem N{\scriptsize EWTON}schen, wonach gleichnamige Elektrizit\"aten sich abstossen, w\"ahrend gleichnamige Massen sich anziehen, entsteht hier durch Auftreten des Faktors: $i^2$. Wir haben n\"amlich die mechanische Masse als Quellpunkt der magnetischen Feldst\"arke und somit als reelle Gr\"osse aufgefasst, w\"ahrend die Elektrizit\"at den Quellpunkt der elektrischen Feldst\"arke darstellt, also imagin\"ar ist.

\chapter{Schlussbemerkungen.}

Es w\"are eine selbst\"andige Leistung der Theorie, wenn es ihr gelingen würde, auch die Belegung als Freiheitsgrade aufzufassen und durch Variation zur elektrischen Belegung und zur mechanischen Masse zu gelangen, wie auch den gegenseitigen Zusammenhang dieser Beiden zu finden. Vorl\"aufig geht die hier gegebene Theorie praktisch nicht bedeutend über die Elektronentheorie hinaus. Es bleibt problematisch, warum die Ladung des Elektrons ebenso wie ihre Masse universelle Konstantent sein müssen. Geradeso bleibt die Wesensverschiedenheit positiver und negativer Elektrizit\"at, wie auch der quantenhafte Charakter der Strahlung\footnote{
Ich m\"ochte hier die Bemerkung einschalten, dass das R\"atsel der positiven Elektrizit\"at und das R\"atsel der Quanten vielleicht eine gemeinsame Wurzel haben. Es besteht n\"amlich sehr wahrscheinlich ein Zusammenhang, welcher die universelle Wirkungsgr\"osse: $h$ aus lauter elektrischen Gr\"ossen auszudrücken erlaubt. Ich schreibe rein empirisch die Gleichung hin:
\begin{equation}
\frac{2 h c}{e^2}~ = ~\frac{\mu_+}{\mu_{-}}
\end{equation}
wo $\mu_{+}$ die Masse des positiven Elektrons, $\mu_{-}$ diejenige des negativen Elektrons, $e$ die Elementarladung in elektrostatischem Masse bezeichnet. Wenn auch diese Gleichung einer numerischen Korrektion bedarf, indem ihr in dieser Form ein Fehler von mehreren Prozenten anhaftet, w\"are es doch sehr unwahrscheinlich, dass zwei dimensionslose Zahlen, welche beide aus universellen Naturkonstanten gebildet sind, aus blossem Zufall miteinander nahe identisch w\"aren.}
 vorl\"aufig unerkl\"art. Doch darf man nicht vergessen, dass man wegen der Kompliziertheit der Aufgabe und auch der mathematischen Behandlung nicht sofort auf schlagende Beweise rechnen kann. Die hier skizzierte Theorie m\"ochte einen Beitrag zum konstruktiven Aufbau der modernen theoretischen Physik liefern, wie er insbesondere durch die Arbeiten Einsteins eingeleitet worden ist. Ihr Wert oder Unwert will eben deshalb nicht nach praktischen positivistisch-oekonomischen Prinzipien beurteilt sein --- da sie keine blosse ``Arbeitshypothese'' sein will. Ihre \"{U}berzeugungskraft --- wenn es sich nicht bloss am meine subjektive T\"auschung handelt --- liegt nicht in ``schlagenden Beweisen,'' sondern in der Folgerichtigkeit und Unwillkürlichkeit ihrer Konstruktion, durch welche sie, die eigentliche Seele der M{\scriptsize AXWELL}schen Gleichungen erfassend, die M{\scriptsize AXWELL}sche Theorie verschmolzen mit der Relativit\"atstheorie naturgem\"ass zu den Elektronen hinführt. In dieser systematischen Einfachheit und Notwendigkeit liegt meiner Ansicht nach ihre \"{U}berlegenheit gegenüber der gew\"ohnlichen Elektronentheorie. Es war mir hier nicht um die ins Einzelne gehende Ausarbeitung, sondern nur um die grossen Umrisse zu tun. Es war mir darum zu tun, mehr deutend, als durchführend einen Weg anzubahnen, dessen Spuren verfolgend vielleicht neue Perspektiven nach den unergründlichen Tiefen der Natur sich er\"offnen werden.

\vskip 3cm

   Es würde auf meine Spekulationen sehr f\"ordernd einwirken, und würde es ebendeshalb hochanschlagen, wenn meine hochgesch\"atzten Fachgenossen, die sich eventuell für diese Gedanken interessieren, ihre gütigen Bemerkungen und Kritiken mir brieflich --- wom\"oglich recommandiert --- mitteilen würden und zwar auf die Adresse : Korn\'el L\'{a}nczos, Assistent an der technischen Hochschule, Institut für Experimentalphysik, Budapest.

\vskip 3cm

Datum dieses Manuskriptes : Oktober 1919.

\newpage

\noindent{\Huge {\bf Nachtrag.}}

~\\

\noindent Ich habe leider erst nachtr\"aglich eine Bemerkung gemacht, welche für die Einheit der Theorie und für die ganze Auffassung von charakteristischer Wichtigkeit zu sein scheint. Sie erbringt zugleich den tats\"achlichen Beweis für die im Texte angedeutete Wesenverwandtschaft der als ``Grundgleichungen'' bezeichneten --- die Quaternionfunktionen definierenden --- Gleichungen mit den C{\scriptsize AUCHY}-R{\scriptsize IEMANN}schen. Letztere lassen sich aus dem Variationsprinzip ableiten, das Minimum für das Integral:
\begin{equation}
\int (X^2 + Y^2) \, dx \, dy 
\end{equation}
zu suchen, wenn ein gegebenes regul\"ares Integrationsgebiet vorliegt und die Werte von:
\begin{equation}
\int  \mathcal{F} \, \nabla \Big( \, \log \frac{1}{R^2} \, \Big) \, dx dy 
\end{equation}
mit:
\begin{equation}
\mathcal{F}=X+iY ~~,~~ \nabla =\frac{\partial ~}{\partial x}+i\frac{\partial ~}{\partial y} ~~,~~ R^2=(x-\xi)^2+(y-\eta)^2 
\end{equation}
für jeden Punkt $( \, \xi, \eta \, )$ des Randes vorgeschrieben sind. In entsprechender Weise folgen die Grundgleichungen in vier Dimensionen aus dem Prinzip:
\begin{equation}
\delta \, \int (X^2 + Y^2 + Z^2 + T^2) \, dx \, dy \, dz \, dt = 0
\end{equation}
bei vorgeschriebenen Randwerten der Gr\"osse:
\begin{equation}
\int \CON{\mathcal{F}} \, \Big( \, \nabla \frac{1}{R^2} \, \Big) \, dx \, dy \, dz \, dt
\end{equation}
($\nabla$ und $\mathcal{F}$, ebenso auch $R$ haben die gewohnte vierdimensionale Bedeutung.)  Dieses selbe Variationsprinzip haben wir aber --- mit komplexen Komponenten für $\mathcal{F}$ und imagin\"arem $t$ --- als ``H{\scriptsize AMILTON}sches Prinzip'' auf die Welt als Ganzes angewendet, (ohne die Grenzbedingungen zu pr\"azisieren,) um zu den Bewegungsgleichungen des Elektrons zu gelangen, wobei also das Integrationsgebiet auch Singularit\"aten enth\"alt. Dieses Prinzip bildet somit die $\emph{universelle Basis}$ der gesamten Theorie, sowohl hinsichtlich der Feldgleichungen, wie auch hinsichtlich der Dynamik. Nur müssten zur exakten Ausführung der Variation die entsprechenden Randbedingungen für das Weltganze gefunden werden.

\end{document}